\documentclass[aps,prd,showpacs,twocolumn,amsmath,nofootinbib]{revtex4-1}
\usepackage{graphicx}
\usepackage{color}
\usepackage{epsfig}
\usepackage{epsf}
\usepackage{dcolumn}
\usepackage{bm}
\usepackage{multirow}
\usepackage{dsfont}
\usepackage{braket}
\usepackage{slashed}
\usepackage[utf8]{inputenc}
\newcommand{\F}{h}

\newcommand{\thetap}{\theta_1}
\begin{document}

\title{Polarization observables in $e^+e^-$ annihilation to 
a baryon-antibaryon pair}

\author{Elisabetta Perotti}
\author{G\"oran F\"aldt}
\author{Andrzej Kupsc}
\author{Stefan Leupold}
\affiliation{Department of Physics and Astronomy, Uppsala University, Box 516, SE-75120 Uppsala, Sweden}
\author{Jiao Jiao Song}
\affiliation{Shandong University, Jinan 250100, People's Republic of China}
\affiliation{Institute of High Energy Physics, Beijing 100049, People's Republic of China}
\date{\today}

\begin{abstract}

Using the helicity formalism of Jacob and Wick we derive spin density matrices of
baryon antibaryon pairs produced in $e^+e^-$ annihilation. We
consider the production of pairs with spins $1/2+\overline{1/2}$,
$1/2+\overline{3/2}$ (+c.c.) and $3/2+\overline{3/2}$. We provide modular
expressions to include chains of weak hadronic two-body decays of the
produced hyperons.  The expressions are suitable for the 
analysis of high statistics data from $J/\psi$ and $\psi(2S)$ decays
at $e^+e^-$ colliders, by fits to the fully differential 
angular distributions of the measured particles. We illustrate the
method by examples, such as the inclusive measurement of the
$e^+e^-\to\psi(2S)\to\Omega^-\bar\Omega^+$ process where one decay
chain $\Omega^-\to\Lambda K^-$ followed by $\Lambda\to p\pi^-$ is
considered. Finally we show that the inclusive 
angular distributions can be used to test 
 spin assignment of the produced baryons.
\end{abstract}

\maketitle
\section{Introduction}
Charmonia are excellent sources of spin entangled hyperon antihyperon
pairs. In particular the states $J/\psi$ or $\psi(2S)$, which carry
$J^{PC}=1^{--}$, are directly produced at electron positron colliders.
They are perfectly suited for precise determination of hyperon decay
parameters and searches for $CP$ symmetry violation in the baryon
sector.

Recent, unexpected observation of polarization in $e^+e^-\to
J/\psi\to\Lambda\bar\Lambda$ at BESIII~\cite{Ablikim:2018zay}
opens up new
  perspectives for such measurements.  The
   polarization allows simultaneous determination of the
  $\Lambda$ and $\bar\Lambda$ decay asymmetries from the events, in
  which all decay products are measured.  Of major importance is the
  new BESIII result for the $\Lambda\to p\pi^-$ asymmetry parameter of
  $\alpha_{-}=0.750\pm0.009\pm0.004$. This decay is used in 
  practically in all experiments involving $\Lambda$ for identification and
  for polarization determination  from the measured
  product of the polarization and the known
  value of the asymmetry parameter. All these
 studies assume the asymmetry parameter of $0.642\pm0.013$, the
 world-average value established in 1978~\cite{Bricman:1978ig}
 and unchanged until the 2018 edition of Review of Particle Physics~\cite{PDG}.
 Therefore the new BESIII value implies that
all published measurements on $\Lambda/\bar\Lambda$ polarization are
$(17\pm3)\%$ too large. This includes e.g. values of decay asymmetries
for weak decays of strange and charmed baryons into final
states including $\Lambda$ such as
$\Xi\to\Lambda\pi$, $\Omega^-\to\Lambda\pi^-$ etc.
   The BESIII
  analysis uses fully differential distributions
  derived in Ref.~\cite{Faldt:2017kgy} using Feynman diagrams formalism.
  Previous $e^+e^-\to J/\psi\to\Lambda\bar\Lambda$
  measurements \cite{Tixier:1988fv,Ablikim:2009ab} used simplified and
  not correct expressions for the amplitudes which precluded such 
  analysis. These expressions were derived using helicity
  formalism of Jacob and Wick~\cite{Jacob:1959at}.
  Therefore, the important tasks is to  repeat the 
  derivation of the angular distributions to make sure the results
  are consistent. In addition the helicity formalism would allow
 to  generalize the angular distributions for the higher spin states.

With a large number of collected $J/\psi$, $(1310.6\pm7.0)\times10^6$,
and $\psi(2S)$, $(448.1\pm2.9)\times10^6$, at the BESIII experiment
\cite{Ablikim:2012cn, Ablikim:2016fal,Ablikim:2012pj} detailed
studies of such systems are now possible\footnote{On Feb. 11th 2019 the BESIII
  Collaboration has announced that $10^{10}$ $J/\psi$ events
were accumulated.}.

Examples of the available
data samples from recent publications are given in
Table~\ref{tab:data}.  The branching fractions, ${\cal B}$, for the listed
decay modes range between $10^{-4}$ and $10^{-3}$ and
the reconstructed data samples are up to $10^6$ events.   In addition,
considering world averages of the ${\cal B}$ values for other $B_1\bar
B_2$ decays, one can anticipate that more modes are accessible with the
collected data sets (Table~\ref{tab:br}).  All of the
published results focus only on the determination of the branching
fractions and the angular distributions of the produced hyperons.

\begin{table*}
\begin{tabular}{p{6cm}|rcr|c}
\hline
decay mode&\multicolumn{2}{l}{events}&&${\cal B} ({\rm units}\ 10^{-4})$\\
\hline
$J/\psi\to\Lambda\bar{\Lambda}$\vphantom{$\int\limits^M$}&$440675$&$\pm$&$670$&${19.43\pm0.03\pm0.33}$\\
$\psi(2S)\to\Lambda\bar{\Lambda}$&$31119$&$\pm$&$ 187$&${3.97\pm0.02\pm0.12} $\\
$J/\psi\to\Sigma^0\bar{\Sigma}^0$&$111026$&$\pm$&$335$&${11.64\pm0.04\pm0.23} $\\
$\psi(2S)\to\Sigma^0\bar{\Sigma}^0$&$6612$&$\pm$&$ 82$&${2.44\pm0.03\pm0.11}$\\
$J/\psi\to\Sigma(1385)^{0}\bar\Sigma(1385)^{0}$&$102762$&$\pm$&$852$&$10.71\pm 0.09 $\\
$J/\psi\to\Xi^{0}\bar\Xi^{0}$&$134846$&$\pm$&$437$&$11.65\pm 0.04 $\\
$\psi(2S)\to\Sigma(1385)^{0}\bar\Sigma(1385)^{0}$&$2214$&$\pm$&$148$&$0.69\pm 0.05 $\\
$\psi(2S)\to\Xi^{0}\bar\Xi^{0}$&$10839$&$\pm$&$123$&$2.73\pm 0.03 $\\
$J/\psi\to\Xi^{-}\bar\Xi^{+}$&$42811$&$\pm$&$231$&$10.40\pm 0.06 $\\
$J/\psi\to\Sigma(1385)^{-}\bar\Sigma(1385)^{+}$&$42595$&$\pm$&$467$&$10.96 \pm 0.12 $\\
$J/\psi\to\Sigma(1385)^{+}\bar\Sigma(1385)^{-}$&$52523$&$\pm$&$596$&$12.58 \pm0.14 $\\
$\psi(2S)\to\Xi^{-}\bar\Xi^{+}$&$5337$&$\pm$&$83$&$2.78\pm 0.05 $\\
$\psi(2S)\to\Sigma(1385)^{-}\bar\Sigma(1385)^{+}$&$1375$&$\pm$&$98$&$0.85\pm 0.06 $\\
$\psi(2S)\to\Sigma(1385)^{+}\bar\Sigma(1385)^{-}$&$1470$&$\pm$&$95$&$0.84\pm 0.05 $\\
\hline
\end{tabular}
\caption[]{Available $B_1\bar B_2$ data 
samples and the branching fractions from recent BESIII publications 
\cite{Ablikim:2017tys,Ablikim:2016sjb,Ablikim:2016iym}.\label{tab:data}}
\end{table*}

\begin{table}
\begin{tabular}{l|rcr}
\hline
decay mode&\multicolumn{3}{c}{${\cal B}({\rm units}\  10^{-4}$)}\\
\hline
$J/\psi\to\Xi(1530)^{-}\bar\Xi^{+}$\vphantom{$\int\limits^M$}&5.9&$\pm$&1.5\\
$J/\psi\to\Xi(1530)^{0}\bar\Xi^{0}$&3.3&$\pm$&1.4\\
$J/\psi\to\Sigma(1385)^{-}\bar\Sigma^{+}$&3.1&$\pm$&0.5\\
$\psi(2S)\to\Omega^-{\bar{\Omega}}^+$&0.47&$\pm$&0.10\\
\hline
\end{tabular}
\caption[]{\label{tab:br} Possible other hyperon antihyperon final states
which can be studied at BESIII. The quoted branching fractions are from
the Particle Data Group~\cite{PDG}. }
\end{table}

The production amplitudes
of such processes are described by a limited set of form factors ---
complex numbers at fixed center-of-mass (CM) energy. For instance, in the 
case of
a spin-1/2 hyperon pair there are just two such form factors.
The angular distribution is described by two real numbers: one
related to the ratio of the absolute values of the form factors and the
other giving the relative phase. In this case, provided that there is
a non-negligible phase between the form factors, one can determine the decay
parameters of the produced hyperons and carry out $CP$
violation tests in the baryon sector.  For the spin-1/2 hyperons with
single-step decay modes
(analogous to $\Lambda$), the formulas provided
in Ref.~\cite{Faldt:2017kgy} could be used directly.  However, to
include other interesting cases the formalism has to be extended for states
where the hyperon antihyperon pair can have a combination of spins
$1/2$ and $3/2$ and for multi-step decay chains.  

Several approaches are suitable to provide 
the amplitude for a
process where the final states carry nonzero spins. We use the helicity formalism originally developed
by Jacob and Wick~\cite{Jacob:1959at}. This formalism had been used in the
past for several hyperon production reactions and decays 
\cite{Berman:1965rc,Luk:1983pe,Tabakin:1985yv,Diehl:1990nu,Guglielmo:1994gb}. However, we did 
not find a general and modular formulation which could be applied
directly to describe high statistics exclusive data, i.e.\ data 
where momenta of all particles are measured for each event.  
For this purpose
fully differential angular distributions are needed, to be used for event
generators and for maximum likelihood fits. It is the purpose of the present paper to document the construction
of such a framework.

We derive spin density matrices for $e^+e^-\to B_1\bar{B}_2$ processes
where the baryon (antibaryon) can have spin $1/2$ or $3/2$. In practice we focus on the cases where all baryons have positive
parity and all antibaryons have negative parity. This fits to the ground state baryons of spin $1/2$ and 
spin $3/2$ \cite{PDG}.  The
presented formalism can be applied to study decays of $J^{PC}=1^{--}$
vector mesons produced in electron positron colliders, such as
$J/\psi$ or $\psi(2S)$, into $B_1\bar{B}_2$ pairs. We will also revise some misleading assumptions and formulas used in the
analyses of  
weak decay chains within this framework.

In order to establish our notation we start with applying the helicity
formalism to the well known case of $1/2+\overline{1/2}$ baryons, then we
proceed to the $1/2+\overline{3/2}$ and $3/2+\overline{3/2}$ cases. We
present a general formalism together with detailed expressions for the spin density matrices
for the production process and for the most important decay modes.

As long as the momentum direction is not flipped, boosts do not change the helicity. 
Therefore in the helicity amplitude method one can disregard the boost part of the Lorentz
group, which allows to obtain angular distributions without using full
expressions for the spinors as required by the Feynman diagram
technique. This is very convenient but comes with a disadvantage: the
energy dependence of the contributing amplitudes cannot be determined
and therefore not even their relative importance. 
Yet for fixed production energy of a two-particle system and for two-body decays of the produced states 
all kinematical variables, i.e.\ all angles, are fully covered by the helicity framework.

We would like to stress again that the basics of our formalism are not new. How to describe in principle the scattering and decays of relativistic particles with spin has been established long time ago. Yet at that time, angular averages were sufficient to account for the available data. Consequently there was no need to provide detailed formulas for the fully differential angular distributions of multi-step decay chains. It is high time to fill this gap in view of the modern high-luminosity experiments, which deliver fully differential data. Only in that way the full potential of presently running and future experiments can be exploited.

The rest of the paper is organized in the following way: In Section \ref{sec:general} we provide the general helicity framework adjusted such that it fits to commonly employed experimental analyses. In Section \ref{sec:prod} we specify to the three production
processes that we are interested in, i.e.\ combinations of spin-1/2 and/or spin-3/2 baryons and antibaryons. Section \ref{sec:decaychains} is devoted to the general discussion of (weak) two-body decay chains. Examples are provided in Section \ref{sec:example}. We have chosen the same examples as considered in Ref.~\cite{Chen:2007zzf}. To facilitate
the matching of theoretical models to experimental results we relate electromagnetic form factors to helicity amplitudes in Section \ref{sec:ff}. Further discussions are provided in Section \ref{sec:discussion}.

\section{General framework}
\label{sec:general}

In general we look at the production of two unstable particles in an
initial scattering reaction. Subsequently the produced particles
decay in one or several steps. The general task is to deduce 
information about the spins and their correlations among the 
involved (unstable) particles. If none of the spins are measured 
directly, this information is encoded in the angular distributions. 
The angles are measured with respect to some axes, which makes it
necessary to define appropriate frames of reference and cartesian 
coordinate systems. 

The production process defines the first coordinate system; see below. 
For the decays it is useful to boost to the rest frame of the mother 
particle. Yet it is helpful to perform rotations before this boost. We 
will be very explicit to motivate and define these rotations.

Following the ideas of \cite{Jacob:1959at,Berman:1965rc} we use the 
helicity formalism. Here the spin quantization axis is not chosen 
along a fixed axis but along the flight direction of the state. The 
advantage is that the helicity does not change when boosting to the 
rest frame of this state. On the other hand, the use of 
angular-momentum (${\bf J}$) conservation for the production and for 
each decay process suggests to single out the $z$-axis, based on the 
convention to use ${\bf J}^2$ and $J_z$ for the characterization of 
states. 

Following this spirit it is useful to spell out how helicity states 
are constructed. To motivate this construction we discuss first 
how one deals with changes of reference frames in experimental analyses.
Afterwards we will describe how to mimic these changes on the theory side.
\paragraph{Experimental procedure:}
 Suppose one has produced a ``mother'' particle that
decays further. One wants to change from the production frame of this state to its rest frame. 
Given the state's three-momentum 
\begin{equation}
  {\bf p}_m = 
p_m \, (\cos\phi_m\sin\theta_m,\sin\phi_m\sin\theta_m,\cos\theta_m)
\end{equation}
and the $z$-axis in the production frame, one possibility would be to 
perform a single rotation that aligns ${\bf p}_m$ with the $z$-axis. 
Subsequently one then boosts to the rest frame of the mother particle. 
The single rotation would be around an axis perpendicular to 
${\bf p}_m$ and $\hat {\bf z}$. Yet when viewed as rotations around 
the coordinate axes this amounts to a succession of three rotations. 
Viewed as active rotations these are 
(a) a rotation around the $z$-axis by $-\phi_m$; (b) a rotation around 
the $y$-axis by $-\theta_m$; (c) a rotation around the $z$-axis by 
$+\phi_m$; see also \cite{Jacob:1959at}. 
In principle, however, the first two rotations are sufficient to align
${\bf p}_m$ with the $z$-axis. In line with the present BESIII analyses 
we follow this two-rotation procedure
in the present work. The rotation matrix for ${\bf p}_m$ is given by
\begin{equation}
\left(
\begin{array}{rrr}
 \cos\theta_m \cos\phi_m & \cos\theta_m \sin\phi_m & -\sin\theta_m \\
 -\sin\phi_m & \cos\phi_m & 0 \\
 \cos\phi_m \sin\theta_m & \sin\theta_m \sin\phi_m & \cos\theta_m \\
\end{array}
\right).\label{eq:hrot}
\end{equation}
This rotation defines in a unique way the helicity reference frame 
for a daughter particle.
In an experimental analysis the  boosts and rotations in Eq~\eqref{eq:hrot}
are applied recursively to all decay products of
a decay chain, thus
defining a set of  helicity variables
to describe an event. 

\paragraph{Matching amplitude:}
To mimic this procedure on the theory side 
we construct helicity states by the inverse procedure, 
following essentially \cite{Berman:1965rc}. A one-particle state with helicity 
$\lambda$ and momentum ${\bf p}=p \, (\cos\phi\sin\theta,
\sin\phi\sin\theta,\cos\theta)$ is constructed from a state 
$\vert p,\lambda \rangle$ that moves along the $z$-direction by
\begin{eqnarray}
  && \vert p,\theta,\phi,\lambda \rangle := 
  R(\phi,\theta,0) \vert p,\lambda \rangle
  \label{eq:ourlinmomstates}
\end{eqnarray}
with \cite{Jacob:1959at}
\begin{eqnarray}
  \label{eq:defR-gen}
  R(\alpha,\beta,\gamma) := e^{-i\alpha J_z} \, e^{-i\beta J_y} \, 
  e^{-i\gamma J_z} \,. 
\end{eqnarray}
Correspondingly a two-particle state in its CM frame is 
given by 
\begin{eqnarray}
  \vert p,\theta,\phi,\lambda_1,\lambda_2 \rangle
  := R(\phi,\theta,0) \vert p,\lambda_1,\lambda_2 \rangle  \,.
  \label{eq:ourlinmomstates2}
\end{eqnarray}

In practice we follow all the steps of \cite{Jacob:1959at}
except for the fact that we use a two-angle rotation procedure as spelled
out in Eq.~\eqref{eq:ourlinmomstates}. 
When constructing \eqref{eq:ourlinmomstates2} the first particle has momentum 
${\bf p} = p \, (\cos\phi\sin\theta,\sin\phi\sin\theta,\cos\theta)$ 
and helicity $\lambda_1$ while the second has momentum $-{\bf p}$ and 
helicity $\lambda_2$. 
The most important consequence of our construction of these two-particle states 
is their projection on angular-momentum eigenstates 
\cite{Berman:1965rc}: 
\begin{eqnarray}\begin{split}
  &\langle J,M,\lambda_1',\lambda_2' \vert 
  \theta, \phi, \lambda_1,\lambda_2 \rangle 
  \\ &=\sqrt{\frac{2J+1}{4\pi}} {\cal D}^{J}_{M,\lambda_1-\lambda_2}(\phi,\theta,0)\, \delta_{\lambda_1\lambda_1'}\delta_{\lambda_2\lambda_2'}
  \label{eq:ang-JM}
\end{split}\end{eqnarray}
where ${\cal D}_{m' m}^j(\alpha,\beta,\gamma) := \braket{jm'|R(\alpha,\beta,\gamma)|jm}$ is the Wigner D-matrix.\footnote{Note that our definition is in line with \cite{Jacob:1959at} but differs from the conventions used in Mathematica \cite{wolfram}. In particular, we have ${\cal D}_{m',m}^j(\alpha,\beta,\gamma)=e^{-i m' \alpha - i m \gamma} {\cal D}_{m',m}^j(0,\beta,0)$
while the built-in ``WignerD'' function of Mathematica satisfies $D_{m',m}^j(\alpha,\beta,\gamma)=e^{i m' \alpha + i m \gamma} D_{m',m}^j(0,\beta,0)$.}

\subsection{Production process}
\label{subsec:prod}

We turn once more to a description of the experimental analysis:
The production process $e^+e^-\to B_1\bar B_2$, 
viewed in the CM frame, defines a 
scattering plane and therefore a coordinate system. The $z$-axis is 
chosen along the line of flight  of the incoming
positron, i.e.\ ${\bf\hat z}={\bf p}_{e^+} = (0,0,p_{\rm in})$, where $p_{\rm in}$ 
denotes the modulus of the momentum of electron and positron in the 
CM frame. The $y$-axis is chosen to be perpendicular to 
the scattering plane. One uses the direction of the baryon $B_1$ to 
define the $y$-axis: 
\begin{eqnarray}
  \label{eq:def-initial-y}
  {\bf\hat y} := \frac{{\bf p}_{e^+} \times {\bf p}_B}%
  {\vert {\bf p}_{e^+} \times {\bf p}_B \vert}  \,.
\end{eqnarray}
Finally the $x$-axis is chosen such that $x$, $y$ and $z$ adhere to the 
right-hand rule. Denoting the scattering angle of $B_1$ 
by $\theta_1$, all this implies 
${\bf p}_B = p_{\rm out} \, (\sin\theta_1,0,\cos\theta_1)$. 
Here $p_{\rm out}$ denotes the modulus of the 
momentum of baryon and antibaryon in the CM frame.
\begin{figure}
\centering
\includegraphics[width=0.48\textwidth]{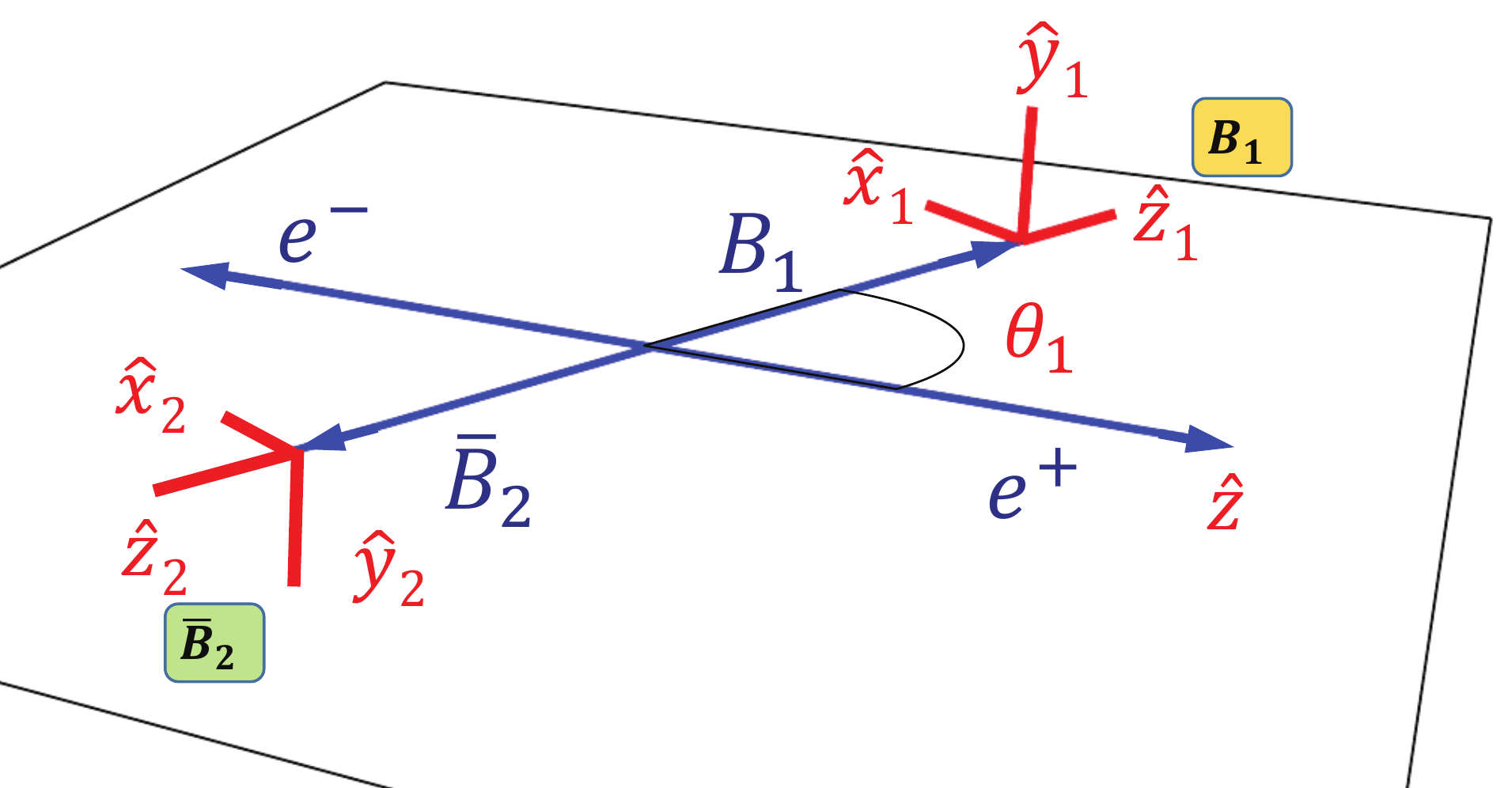}
\caption[]{(color online) Orientation of the axes in baryon $B_1$ and
  antibaryon $\bar B_2$ helicity frames.}
  \label{fig:axes}
\end{figure}
  
With the above definition of the CM coordinate system,
the $y$ axis of
the helicity frame of the baryon $B_1$, ${\bf\hat y}_1$ in Fig.~\ref{fig:axes}, is the same as ${\bf\hat y}$ in
Eq.~\eqref{eq:def-initial-y}.
Therefore, for the helicity rotation matrix
Eq.~\eqref{eq:hrot} one uses $\theta_m =\theta_1$ and $\phi_m =0$.
Correspondingly, to transform to the helicity frame of the antibaryon $\bar B_2$
one chooses $\phi_m=\pi$ and $\theta_m= \pi-\theta_1$. In this way 
the $y$-axis, ${\bf\hat y}_2$, is equal $-{\bf\hat y}$. The
$y$- and $z$-axes of the helicity frames of the baryon $B_1$ and the antibaryon
$\bar B_2$ have opposite directions while it is the same direction for the $x$-axis as shown in Fig.~\ref{fig:axes}.

Now we turn to the theoretical construction that goes along with 
the experimental analysis:
Let $\lambda$ denote the initial helicity of the positron. 
Neglecting the mass of the electron and working within the 
one-photon approximation this implies that the helicity of the 
electron is $-\lambda$ since the photon only couples 
right-handed particles to left-handed antiparticles and vice versa. 
Since $\lambda$ can take the values $\pm 1/2$, then the helicity 
difference $k:=\lambda-(-\lambda) = \pm 1$. 

For unpolarized initial states one sums over $\lambda$ or 
equivalently over $k=2\lambda$. The density matrix for the production is 
proportional to
\begin{eqnarray}
  && \rho^{\lambda_1,\lambda_2;\lambda_1',\lambda_2'}_{B_1\bar B_2} \nonumber 
  \propto \sum\limits_{k=\pm 1} {}_{\rm o}\!\langle \thetap, 0, 
  \lambda_1,\lambda_2 \vert S \vert 0,0,\lambda,-\lambda\rangle_{\rm i} \; 
  \nonumber \\ && \phantom{\sim} \times 
  _{\rm i}\!\!\langle 0,0,\lambda,-\lambda \vert S^\dagger \vert 
  \thetap, 0, \lambda_1',\lambda_2' \rangle_{\rm o}  \, ,
  \label{eq:prod-dens}  
\end{eqnarray}
where we use the $\langle$bra$\vert$, $\vert$ket$\rangle$  
notation with index $\rm i$ and $\rm o$ to denote
in and out states, respectively. 
Now we  evaluate the transition operator $S$:
\begin{eqnarray}
  && {}_{\rm o}\!\langle \thetap, 0, \lambda_1,\lambda_2 \vert S \vert 0,0,
\lambda,-\lambda\rangle_{\rm i}   \nonumber \\
  && = \sum\limits_{J,M} {}_{\rm o}\!\langle \thetap, 0, \lambda_1,\lambda_2 \vert JM,\lambda_1,\lambda_2 \rangle_{\rm o}
  \nonumber \\ && \times 
  {}_{\rm o}\!\langle JM,\lambda_1,\lambda_2 \vert S \vert JM,\lambda,-\lambda \rangle_{\rm i} 
  \nonumber \\ && \times
  {}_{\rm i}\!\langle JM,\lambda,-\lambda \vert 0,0,\lambda,-\lambda\rangle_{\rm i}   \,.
  \label{eq:evalS1}
\end{eqnarray}
We have to evaluate three matrix elements. The first and the third bring in Wigner functions. 
The general formula is given in Eq.~\eqref{eq:ang-JM}.
For the transition amplitude one finds in the one-photon approximation
\begin{eqnarray}
  && {}_{\rm o}\!\langle JM,\lambda_1,\lambda_2 \vert S \vert JM,\lambda,-\lambda \rangle_{\rm i} \nonumber \\ 
  && \approx {}_{\rm o}\!\langle JM,\lambda_1,\lambda_2 \vert S_{\gamma^* \to {\rm out}} \, 
  S_{{\rm in}\to \gamma^*} \vert JM,\lambda,-\lambda \rangle_{\rm i} \nonumber \\ 
  && 
  = \delta_{J,1}  \,  \, A_{\lambda_1,\lambda_2} \, A^{{\rm in}}_{\lambda,-\lambda}   \,.
  \label{eq:JM-S}
\end{eqnarray}
 Here $A_{\lambda_1,\lambda_2}$ denotes the transition amplitude between 
helicity states. 
Only transitions fulfilling the inequality
\begin{equation}
  |\lambda_1-\lambda_2|\le J = 1
  \label{eq:limit}
\end{equation} 
are different from zero. For a parity conserving process the
amplitudes between opposite helicity states are related:
\begin{equation}
A_{\lambda_1,\lambda_2}=\eta_1\eta_2\eta(-1)^{J-s_1-s_2}A_{-\lambda_1,-\lambda_2}\label{eq:parity}\,,
\end{equation} 
where $\eta$ is the parity of the initial state, $\eta_1$ and $\eta_2$ are the parities of the final state particles.  
Moreover parity symmetry of QED implies $A^{{\rm in}}_{-\lambda,\lambda} = A^{{\rm in}}_{\lambda,-\lambda}$ for the initial $e^+e^- \to \gamma^*$ 
production amplitude $A^{{\rm in}}$. Here we are not interested in the 
$p_{\rm in}$ dependence of the reaction and therefore we can drop $A^{{\rm in}}$.
One finds
\begin{eqnarray}
  && {}_{\rm o}\!\langle \thetap, 0, \lambda_1,\lambda_2 \vert S \vert 0,0,\lambda,-\lambda\rangle_{\rm i}   \nonumber \\
  && \propto \sum\limits_M [{\cal D}_{M,\lambda_1-\lambda_2}^{1}(0,\thetap,0)]^* \, A_{\lambda_1,\lambda_2} \, {\cal D}_{M,k}^{1}(0,0,0) \nonumber \\ 
  && = [{\cal D}_{k,\lambda_1-\lambda_2}^{1}(0,\thetap,0)]^* \, A_{\lambda_1,\lambda_2}  \,. 
  \label{eq:evalS2}  
\end{eqnarray}
We obtain for the production density matrix: 
\begin{eqnarray}
  \rho^{\lambda_1,\lambda_2;\lambda_1',\lambda_2'}_{B_1\bar B_2} \propto A_{\lambda_1,\lambda_2} \, A^*_{\lambda'_1,\lambda'_2} \, \rho_1^{\lambda_1-\lambda_2,\lambda'_1-\lambda'_2}(\thetap\!) 
  \label{eqn:amp}  
\end{eqnarray}
with 
\begin{eqnarray}
  \rho_1^{i,j}(\theta) := 
  \sum\limits_{k=\pm 1}
  {\cal D}_{k,i}^{1*}(0,\theta,0) \, {\cal D}^1_{k,j}(0,\theta,0) \,.
  \label{eq:defrho0init}  
\end{eqnarray}
The explicit form of the reduced density matrix $\rho_1$ is given by
\begin{equation}
\rho_1(\theta)=\left(
\begin{array}{ccc}
 \frac{1+\cos^2\!\theta}{2} & 
-\frac{\cos\theta\sin\theta}{\sqrt{2}} & \frac{\sin ^2\!\theta}{2} \\[0.5em]
-\frac{\cos\theta\sin\theta}{\sqrt{2}} & 
\sin ^2\!\theta & \frac{\cos\theta\sin\theta}{\sqrt{2}} \\[0.5em]
\frac{\sin ^2\!\theta}{2} & \frac{\cos \theta \sin \theta}{\sqrt{2}}  &
\frac{1+\cos^2\!\theta}{2}  \\
\end{array}
\right)  \,.
\label{eq:explicit-init}
\end{equation}

We note in passing that here one could also rotate to a frame where the baryons do not lie in the $x$-$z$ plane, i.e.\ where they have a 
non-vanishing value of $\phi$. This would {\em not} change the density matrix because of the following relation:
\begin{equation}
  {\cal D}_{k,i}^{1*}(0,\theta,0) \, {\cal D}^1_{k,j}(0,\theta,0) = 
  {\cal D}_{k,i}^{1*}(\phi,\theta,0) \, {\cal D}^1_{k,j}(\phi,\theta,0)  \,.
\end{equation}
{This also points to the core difference with all previous helicity amplitude
  calculations of the $e^+e^-\to B_1\bar B_2$ process starting from Ref.~\cite{Tixier:1988fv}. All they
  obtain   the initial $\rho_1$ density matrix which
  is dependent on $\phi$. This is an unphysical result for
  transversely unpolarized electron and positron beams
  due to the rotation symmetry with respect to the ${\bf\hat z}$
  axis. The
  unwanted $\phi$
  dependence is then eliminated  by an arbitrary integration over
  the $\phi$ variable. The result is a diagonal density matrix
  and all interference terms between heicity amplitudes
  of the produced baryons cancel. We can reproduce 
  all results from Refs.~\cite{Tixier:1988fv,Chen:2007zzf}
  by using the diagonal part of $\rho_1$ from Eq.~\eqref{eq:explicit-init}:
  ${\rm diag}((1+\cos^2\!\theta)/2,\sin^2\!\theta,(1+\cos^2\!\theta)/2)$.
}

Finally we note that for the case where $B_1$ and $B_2$ are of the same
type and in the one-photon approximation, charge conjugation provides
the following (schematic) relation:
$\langle \gamma^* \vert S \vert B_1, \bar B_2 \rangle$ = 
$\langle \gamma^* \vert S \vert B_2, \bar B_1 \rangle$. The minus sign 
emerging from the virtual photon is compensated by the reordering of 
the two (anti-commuting) fermions from $\vert \bar B_1 , B_2 \rangle$ to
$\vert B_2, \bar B_1 \rangle$.

\subsection{Baryon spin density matrices}
\label{sec:Bsdm}

The most general spin density matrix for a spin-1/2 particle has
the following form:
\begin{equation}
\rho_{1/2}=\frac{1}{2}\left(
\begin{array}{cc}
I_0+I_z&I_x-iI_y\\
I_x+iI_y&I_0-I_z,
\end{array}
\right)
\end{equation}
or expressed  in a compact way:
\begin{equation}
\rho_{1/2}=\frac{1}{2}\sum_\mu I_\mu\sigma_\mu\,,
\label{eqn:dens12}
\end{equation}
where $\mu=0,x,y,z$; $\sigma_x,\sigma_y,\sigma_z$ are the Pauli
matrices and $\sigma_0$ is the identity $2\times 2$ matrix.  $I_0$ is
the cross section term and ${\bf I}$ is a three vector ${\bf
  I}=I_0\cdot {\bf P}$, where ${\bf P}$ is the polarization vector for
the fermion. For some formulas we also use notation 
with a numeric index: $\mu=0,1,2,3$.

The density matrix of a spin-3/2 particle can be written in terms of sixteen Hermitian
$4\times 4$ matrices $Q_{\mu}$ with $\mu=0,...,15$ as described in
Ref.~\cite{Doncel:1972ez}. The explicit expression
for these matrices is given  in 
Appendix~\ref{sec:mat32}. The general density matrix for a single spin-3/2 particle
can be expressed as 
\begin{equation}
\rho_{3/2}=\sum_{\mu=0}^{15}r_\mu Q_\mu\, ,
\label{eqn:dens32don}
\end{equation}
where $r_0$ is the cross section term, $Q_0$ is $\frac{1}{4}\mathds{1}_4$ where $\mathds{1}_4$ is the  $4\times 4$ identity
matrix and $r_\mu$ are real numbers.

\section{Specific production processes}
\label{sec:prod}

\subsection{Two spin-$\frac{1}{2}$ baryons}
\label{sectionI}

It is well known how the spin density matrices look like for a reaction $e^+e^-\to B_1\bar{B}_2$ where both produced 
particles have spin 1/2. The results were 
obtained using different approaches
\cite{Dubnickova:1992ii,Czyz:2007wi,TomasiGustafsson:2005kc,Faldt:2013gka,Faldt:2016qee,Faldt:2017kgy}. Here we reproduce the result 
using the helicity method. We focus on the case where the baryon has positive parity $\eta_1=1$ and the antibaryon 
negative parity $\eta_2=-1$. This fits to the production of a pair of ground-state hyperons.
In general only two out of four possible helicity transitions are
independent. Using $\eta_1\eta_2 =-1$ for the baryon
antibaryon pair one can set
$A_{1/2,1/2}=A_{-1/2,-1/2}=:\F_1$ and $A_{1/2,-1/2}=A_{-1/2,1/2}=:\F_2$.
The transition amplitude matrix is
\begin{equation}
\left(
\begin{array}{cc}
 {\F_1} & {\F_2} \\
 {\F_2} & {\F_1} \\
\end{array}
\right)\, .
\end{equation}

The spin density matrix for a two-particle $1/2+\overline{1/2}$ system can be expressed
in terms of a set of $4\times 4 $ matrices obtained
from the outer product, $\otimes$, of $\sigma_\mu$ and ${\sigma}_{\bar\nu}$ \cite{Tabakin:1985yv}:
\begin{equation}
  \rho_{B_1,\bar B_2}=\frac{1}{4}\sum_{\mu,\bar\nu=0}^3C_{\mu\bar\nu}(\thetap\!)\,
  \sigma_\mu^{B_1}\otimes
      {\sigma}_{\bar\nu}^{\bar B_2},
\label{eqn:sig12}
\end{equation}
{where $\sigma_\mu^{B}$ with $\mu=0,1,2,3,4$ represent spin-$1/2$
  base  matrices
  for a  baryon $B$ in the rest frame. The $2\times 2$ matrices
 are $\sigma_0^{B}=\mathds{1}_2$, 
$\sigma_1^{B}=\sigma_x$, $\sigma_2^{B}=\sigma_y$ and $\sigma_3^{B}=\sigma_z$.
In particular the spin matrices $\sigma_\mu^{B_1}$ and
${\sigma}_{\bar\nu}^{\bar B_2}$ are given in the helicity frames of
the baryons $B_1$ and $\bar B_2$, respectively. The axes of the frames
are defined in Fig.~\ref{fig:axes} and denoted by
${\bf\hat x}_1,{\bf\hat y}_1,{\bf \hat z}_1$ and ${\bf\hat x}_2,{\bf\hat y}_2,{\bf\hat z}_2$.
  The real coefficients $C_{\mu\bar\nu}$
  are functions of the scattering angle $\thetap$ of $B_1$.
}

Suppose one is not interested in the absolute size of the cross section but only in the 
(not normalized) angular distributions. For their description we do not need all information
contained in the two complex form factors $\F_1$ and $\F_2$. Instead we can use just two real
parameters: First, $\alpha_\psi$ as defined below and, second, the relative phase between the form
factors $\Delta\Phi=\arg(\F_1/\F_2)$, i.e.\ we disregard the normalization and 
the overall phase.  More specifically without any loss of generality we take $\F_1$ as
real and set $\F_1=\sqrt{1-\alpha_\psi}/\sqrt{2}$ and
$\F_2=\sqrt{1+\alpha_\psi}\exp(-i\Delta\Phi)$.
Only 8 coefficients $C_{\mu\bar\nu}$ are non-zero and they are given by
\begin{eqnarray}
C_{00}&=&2(1+\alpha_\psi\cos^2\!\thetap\!)   \,, \nonumber\\
C_{0 2}&=&2\sqrt{1-\alpha_\psi^2}\sin\thetap\cos\thetap\sin(\Delta\Phi)   \,, \nonumber\\
C_{1 1}&=&2\sin^2\!\thetap   \,, \nonumber\\
C_{1 3}&=&2\sqrt{1-\alpha_\psi^2}\sin\thetap\cos\thetap\cos(\Delta\Phi)   \,, \nonumber\\
C_{20}&=&-C_{0 2}   \,,\label{eqn:c1212} \\
C_{2 2}&=&\alpha_\psi C_{11}   \,, \nonumber\\
C_{3 1}&=&-C_{1 3}   \,, \nonumber\\
C_{3 3}&=&-2(\alpha_\psi+\cos^2\!\thetap\!)\, .\nonumber
\end{eqnarray}

For the case when the antibaryon $\bar B_2$
is not measured (the decay products are not registered),
the corresponding {\it inclusive} density matrix can be obtained
by taking the trace of the formula in Eq.~\eqref{eqn:sig12} with respect 
to the spin variables of $\bar B_2$. The result is
\begin{equation}
\rho_{B_1}=\frac{1}{2}\sum_{\mu}C_{\mu0}\, \sigma_\mu^{B_1}\, ,
\label{eqn:sig12a}
\end{equation}
where
\begin{eqnarray}
C_{00}&=&I_{0}=2(1+\alpha_\psi\cos^2\!\thetap\!)   \,, \nonumber \\
C_{20}&=&I_y=-2\sqrt{1-\alpha_\psi^2}\sin\thetap\cos\thetap\sin(\Delta\Phi) \,. \phantom{mm}
\end{eqnarray}
 
If the produced spin-1/2 baryon is a hyperon decaying weakly, one
can determine the polarization of $B_1$ in the $e^+e^-\to
B_1\bar{B}_2$ production process from the angular distributions of the
decay products.  The most common case is a weak decay into a spin-1/2
fermion and a pseudoscalar ({\it e.g.} $\Lambda\to p \pi^-$).
For the case of a one-step process, when the decay product is stable
and its polarization is not measured, the final angular
distribution is given by:
\begin{equation}
{d\sigma} \propto (I_0+\alpha_1  I_y\sin\theta_\mathrm{p} \sin\phi_\mathrm{p}){d\Omega_\mathrm{p}}\, ,
\end{equation}
where $\alpha_1$ is the decay asymmetry parameter for the
corresponding weak decay mode of $B_1$.

\subsection{Spin $\frac{1}{2}$ and spin $\frac{3}{2}$ baryon}
\label{section1213}

To be specific we consider $e^+e^-\to B_1\bar{B}_2$ where $B_1$
has spin 1/2 and $\bar{B}_2$ spin 3/2.
We focus on the case where the baryon has positive parity, $\eta_1=1$, and the antibaryon 
negative parity, $\eta_2=-1$. This fits to the production of ground-state hyperons with the respective spins.
In general only three out of eight transition amplitudes are independent: Parity symmetry of the 
production process relates the amplitudes pairwise. In addition, the one-photon approximation does not allow
for the helicity combination where $\vert \lambda_1 - \lambda_2 \vert =2$ on account of Eq.~\eqref{eq:limit}. 

Again we have $\eta_1\eta_2 = -1$
for the baryon antibaryon 
pair so that $A_{\lambda_1,\lambda_2}=-A_{-\lambda_1,-\lambda_2}$ follows from Eq.~\eqref{eq:parity}. 
For simplicity we introduce $A_{1/2,1/2}=-A_{-1/2,-1/2}=:\F_1$,
$A_{1/2,-1/2}=-A_{-1/2,1/2}=:\F_2$  and 
 $A_{1/2,3/2}=-A_{-1/2,-3/2}=:\F_3$. In the one-photon approximation the remaining amplitudes vanish: 
$A_{-1/2,3/2}=A_{1/2,-3/2}=0$.
Therefore the transition amplitude can be expressed as:
\begin{equation}
\left(
\begin{array}{cccc}
 {\F_3} &  {\F_1} & {\F_2} & 0 \\
 0 & -{\F_2} & -{\F_1} & -{\F_3} \\
\end{array}
\right)\, .
\end{equation}

The density matrix for the $1/2+\overline{3/2}$ system can be expressed in terms of a set of $8\times 8 $ matrices
obtained from the outer product of $\sigma_\mu$ and 
${Q}_{\bar\nu}$:
\begin{equation}
\rho_{B_1,\bar B_2}=\frac{1}{2}
\sum_{\mu=0}^3\sum_{\bar\nu=0}^{15}C_{\mu\bar\nu}(\thetap\!)\,\sigma_\mu^{B_1}\otimes Q_{\bar\nu}^{\bar B_2}\, ,
\label{eqn:dens12x32}
\end{equation}
{where the spin matrices $\sigma_\mu^{B_1}$ and
${Q}_{\bar\nu}^{\bar B_2}$ are given in the helicity frames of
  the baryons $B_1$ and $\bar B_2$, respectively. In principle
  there are $4\times 16$ real functions
$C_{\mu\bar\nu}(\thetap\!)$, but only 30 are non-zero.} 
Here we just give the expressions for the inclusive spin density matrices
for the $1/2$ and the $\overline{3/2}$ baryon, respectively.

The inclusive density matrix for the spin-1/2 baryon $B_1$ is obtained by 
taking the trace of the formula in Eq.~\eqref{eqn:dens12x32} with respect 
to the spin variables of the antibaryon $\bar B_2$. One obtains the general
form \eqref{eqn:dens12} with entries
\begin{eqnarray}
I_{0}&= & 2|\F_1|^2 \sin^2\thetap+ (1+\cos^2\thetap\!) (|\F_2|^2+|\F_3|^2)   \,, \nonumber \\
I_y&= & 2\sqrt{2}\ {\Im}(\F_1\F_2^*)\sin\thetap\cos\thetap    \,, \\
I_x&=&I_z=0 \,. \nonumber 
\end{eqnarray}

The corresponding inclusive spin density 
matrix obtained for the baryon $\bar B_2$
can be expressed as
\begin{equation}
\rho_{\overline{3/2}}(\thetap\!)=\left(
\begin{array}{cccc}
 m_{11} & c_{12} & c_{13} & 0 \\
 c_{12}^* & m_{22} & i m_{23} & c_{13}^* \\
 c_{13}^* & -i m_{23} & m_{22} & -c_{12}^* \\
 0 & c_{13} & -c_{12} & m_{11} \\
\end{array}
\right)\, ,
\label{eq:dens32}
\end{equation}
where $m_{11}$, $m_{22}$ and  $m_{23}$ are real
while  $c_{12}$ and $c_{13}$ are complex functions of
the scattering angle $\thetap$.
These elements of the spin density matrix are
\begin{eqnarray}
m_{11}&=& \frac{1+\cos^2\!\thetap }{2} |\F_3|^2   \,, \nonumber  \\
m_{22}&=&|\F_1|^2 \sin^2\!\thetap+\frac{1+\cos^2\!\thetap}{2}  |\F_2|^2   \,,  \nonumber \\
m_{23}&=&{\sqrt{2}}\ {\Im({\F_2}\F_1^*) \cos\thetap \sin\thetap}    \,, \\
c_{12}&=& \frac{\F_3 \F_1^* \cos\thetap \sin\thetap}{\sqrt{2}}    \,,  \nonumber \\
c_{13}&=& \frac{1}{2}{\F_3}\F_2^*\sin^2\!\thetap \,.   \nonumber 
\end{eqnarray}

The density matrix $\rho_{\overline{3/2}}$ can be also written in terms of the polarization parameters introduced in Eq.~\eqref{eqn:dens32don}. Since we are considering a parity conserving process it turns out that only seven parameters 
are non-zero: $r_{0}$, $r_{1}$, $r_6$, $r_7$, $r_8$, 
$r_{10}$ and  $r_{11}$. This fits to the previous seven parameters: $m_{11}$, $m_{22}$, $m_{23}$ and real and imaginary part of 
$c_{12}$ and $c_{13}$. 
The former are expressed as functions of the scattering angle
$\thetap$ in the following way: 
\begin{eqnarray}
r_0&=&(\cos ^2\!\thetap+1)(|\F_2|^2+|\F_3|^2)+2|\F_1|^2\sin^2\!\thetap  \,, \nonumber\\
r_{1}&=&2\sin\!2\thetap \frac{2\Im(\F_1\F_2^*)
+\sqrt{3} \Im(\F_1 \F_3^*)}{\sqrt{30}}  \,, \nonumber \\
r_6&=&-\frac{2\sin ^2\!\thetap
|\F_1| ^2+(|\F_2|^2-|\F_3|^2)(\cos^2\!\thetap +1)}{ \sqrt{3}}  \,, \nonumber\\
r_7&=&\sqrt{2} \sin\!2 \thetap\frac{\Re(\F_1\F_3^*)}{\sqrt{3}}   \,, \\
r_8&=&2\sin ^2\!\thetap \frac{\Re(\F_2\F_3^*)}{\sqrt{3}}  \,, \nonumber\\
r_{10}&=&2\sin ^2\!\thetap \frac{\Im(\F_2\F_3^*)}{\sqrt{3}}  \,, \nonumber \\
r_{11}&=&2 \sin\!2 \thetap \frac{\Im\left(\sqrt{3}{\F_2} \F_1^*+
\F_1 \F_3^*\right)}{\sqrt{15}}\, .\nonumber
\end{eqnarray}

\subsection{Two spin-$\frac{3}{2}$ baryons}
\label{section33}

We focus again on the case where the baryon has positive parity $\eta_1=1$ and the antibaryon 
negative parity $\eta_2=-1$. This fits to the production of ground-state hyperons with spin 3/2.
Actually all such ground-state hyperons are distinct from each other by strangeness or electric charge. Thus we focus on the case
where the produced antibaryon is the antiparticle of the produced baryon (and not an arbitrary spin-3/2 state). This allows 
to involve arguments from charge conjugation invariance.

For $e^+e^-\to B_1\bar{B}_2$, where both $B_1$
and $\bar{B}_2$ are spin-3/2 particles, only 4 out of 16 amplitudes
are independent. From Eq.~\eqref{eq:parity} it follows that $A_{\lambda_1,\lambda_2}=A_{-\lambda_1,-\lambda_2}$. We only need to consider
$A_{1/2,-1/2}=A_{-1/2,1/2}=:\F_2$, $A_{1/2,1/2}=A_{-1/2,-1/2}=:\F_1$, 
$A_{-3/2,-1/2}=A_{3/2,1/2}=:\F_3$, and $A_{-3/2,-3/2}=A_{3/2,3/2}=:\F_4$. 
Due to Eq.~\eqref{eq:limit} expressing the constraint for the spin projection values of the 
initial state (one-photon approximation) the following amplitudes vanish: $A_{-1/2,3/2}=A_{1/2,-3/2}=0$. 
Moreover $A_{-1/2,-3/2}=A_{1/2,3/2}=\F_3$ due to charge conjugation invariance.
Thus the transition amplitude is given by
\begin{equation}
\left(
\begin{array}{cccc}
 {\F_4} & {\F_3} & 0     & 0 \\
 {\F_3} & {\F_1} & {\F_2} & 0 \\
 0     & {\F_2} & {\F_1} & {\F_3} \\
 0     & 0     & {\F_3} & {\F_4} 
\end{array}
\right)\, .
\end{equation}

The density matrix for the $3/2+\overline{3/2}$ system can be expressed
in terms of a set of $16\times 16 $ matrices constructed from
the outer product of $Q_{\mu}$ and  $Q_{\bar\nu}$:
\begin{equation}
\rho_{B_1,\bar B_2}=
\sum_{\mu=0}^{15}\sum_{\bar\nu=0}^{15}C_{\mu\bar\nu}\, Q_{\mu}^{B_1}\otimes Q_{\bar\nu}^{\bar B_2}\, ,
\label{eqn:dens32x32}
\end{equation}
where $C_{\mu\bar\nu}(\thetap\!)$ is a set of 256 real functions
of $\thetap$ of which 140
are zero.

If the antibaryon is not registered the inclusive density matrix of 
the spin-3/2 baryon $B_1$ is again given by Eq.~\eqref{eq:dens32}.
In this case, the elements are
\begin{eqnarray}
m_{11}&=& \frac{ 1+\cos^2\!\thetap}{2} {|\F_3|^2}+{|\F_4|^2}\sin^2\!\thetap  \,,  \nonumber  \\
m_{22}&=&\frac{ 1+\cos^2\!\thetap}{2} \left(\left| {\F_2}\right|^2+\left| {\F_3}\right| ^2\right)+{|\F_1|^2}\sin^2\!\thetap  \,,  \nonumber  \\
m_{23}&=&{\sqrt{2}}\ \Im({\F_1} \F_2^*) \cos  \thetap  \sin  \thetap   \,,  \\
c_{12}&=&\frac{1}{\sqrt{2}}\left(\F_4 \F_3^* - \F_3 \F_1^* \right)\cos  \thetap\sin  \thetap   \,,   \nonumber \\
c_{13}&=& \frac{1}{2} {\F_3} \F_2^*\sin ^2\! \thetap \,.   \nonumber 
\end{eqnarray}
The angular distribution is given by the trace of the density matrix:
\begin{equation}
\frac{d\sigma}{d\cos\thetap}\propto 2(m_{11}+m_{22})\, .
\end{equation}
Defining 
\begin{equation}
  \alpha_\psi=\frac{\left| {\F_2}\right| ^2-2 (\left|{\F_1}\right| ^2-\left| {\F_3}\right| ^2+ \left| {\F_4}\right| ^2)}
  {\left| {\F_2}\right| ^2+2 (\left|{\F_1}\right| ^2+\left| {\F_3}\right| ^2+ \left| {\F_4}\right| ^2)} \,,
  \label{eq:aphapsi32}
\end{equation}
it can be written as $1+\alpha_\psi\cos^2\!\thetap$.
Using Eq.~\eqref{eqn:dens32don} an alternative representation for the inclusive 
density matrix for the spin-3/2 baryon 
is given by the following seven real $r_\mu=C_{\mu 0}$ coefficients
(the remaining nine are zero):
\begin{eqnarray}
r_0&=&\left[\left| {\F_2}\right| ^2\!+\!2 (\left|{\F_1}\right| ^2\!+\!\left| {\F_3}\right| ^2\!+\! \left| {\F_4}\right| ^2)\right](1\!+\!\alpha_\psi\cos^2\!\thetap\!)   \,, \nonumber
\\ 
r_1&=&2\sin\!2\thetap \frac{2\Im(\F_2\F_1^*)
+\sqrt{3} \Im(\F_3 (\F_1^*+\F_4^*))}{\sqrt{30}}   \,, \nonumber \\
r_{6}&=&-\frac{2\sin ^2\!\thetap
\left(\left|\F_1\right| ^2-|\F_4|^2\right)+|\F_2|^2(\cos^2\!\thetap +1)}{ \sqrt{3}}   \,, \nonumber\\
r_7&=&\sqrt{2} \sin\!2 \thetap\frac{\Re(\F_3^*(\F_4-\F_1))}{\sqrt{3}}   \,, \label{eqn:r32} \\
r_8&=&2\sin ^2\!\thetap \frac{\Re(\F_2\F_3^*)}{\sqrt{3}}   \,, \nonumber\\
r_{10}&=&2\sin ^2\!\thetap \frac{\Im(\F_2\F_3^*)}{\sqrt{3}}   \,, \nonumber \\
r_{11}&=&2 \sin\!2 \thetap \frac{\Im\left(\sqrt{3}{\F_1} \F_2^*+
\F_3 (\F_1^*+\F_4^*)\right)}{\sqrt{15}}.\nonumber
\end{eqnarray}
The corresponding coefficients for the inclusive 
density matrix of the antibaryon are the same, provided 
one uses the scattering angle of the antibaryon, i.e.\ $\thetap\to \pi-\thetap$.
 
\section{Decay chains}
\label{sec:decaychains}

The density matrices of the produced hyperons can be used to derive
the angular distributions of the particles produced in the subsequent
decays.  When considering multi-step decay processes, also the density matrices of the
intermediate states are needed.  Moreover one should keep track
of the spin correlations for the initial $B_1\bar B_2$ pair.
We propose a general modular method to obtain the distributions in a
systematic way. Since the joined production density matrices of
Eqs.~\eqref{eqn:sig12}, \eqref{eqn:dens12x32} and \eqref{eqn:dens32x32}
are expressed as outer products of the basis matrices $\sigma_\mu$ and $Q_\mu$,
 it is enough to know how the latter individually transform under a decay process. 

We consider two weak decay modes, which cover most of the relevant 
cases\footnote{More cases are discussed e.g. in Ref.~\cite{Faldt:2017yqt}.}:
1) spin-$3/2^+$ hyperon decaying into spin-$1/2^+$ hyperon and 
pseudoscalar, 2) spin-$1/2^+$ hyperon decaying into spin-$1/2^+$
hyperon and  pseudoscalar.  If we neglect the widths of the
initial and final particles, the CM momentum of the decay
particles is fixed. The angular distribution is specified by two
spherical angles $\theta$ and $\phi$, which give the direction of the
final baryon in the helicity frame of the initial hyperon.  The spin
configuration of the final system is fully specified by the spin density
matrix of the final baryon, which has spin $1/2$ in both cases, since
the accompanying particle is a pseudoscalar meson. Let us start considering a decay of type 1). The aim is to relate the basis matrices of the
mother hyperon $Q_\mu$ to those of the daughter
baryon $\sigma_\nu^d$. In other words one has to find the {\it transition matrix}
$b_{\mu\nu}$  such that:
\begin{equation}
Q_\mu\to\sum_{\nu=0}^3 b_{\mu\nu}\sigma_\nu^d\, .
\end{equation}
 The $16\times 4$ $b_{\mu\nu}$ matrix 
depends only on the final baryon $\theta$ and $\phi$ angles, and
on the decay parameters of the considered decay mode. 
If the initial particle density matrix
is given by Eq.~\eqref{eqn:dens32don} then the final baryon 
density matrix is:
\begin{equation} \label{eq:rhoDaughter}
\rho_{1/2}^d=\sum_{\mu=0}^{15} \sum_{\nu=0}^3 r_\mu b_{\mu\nu}\sigma_\nu^d.
\end{equation}
The differential
cross section is simply obtained by taking the trace of $\rho_{1/2}^d$:
\begin{equation}
\mathrm{Tr}\, \rho_{1/2}^d=2\sum_{\mu=0}^{15} r_\mu b_{\mu,0}.
\end{equation}

Let us now consider a decay of type 2). Similarly we introduce 
a $4\times 4$ matrix $a_{\mu\nu}$ which allows us to express the
$\sigma_\mu$ matrices in the mother helicity frame in terms of $\sigma_\nu^d$
matrices in the daughter helicity frame:  
\begin{equation}
\sigma_\mu\to\sum_{\nu=0}^3 a_{\mu\nu}\sigma_\nu^d\, . \label{eqn:decay12p}
\end{equation}
The decay matrices $a_{\mu\nu}$ and $b_{\mu\nu}$ introduced above 
allow to keep track of the spin correlation between the decay 
products of the $B_1$ and $\bar B_2$ decays chains.

In the following example we start from the two-particle $1/2+\overline{3/2}$
density matrix given by Eq.~\eqref{eqn:dens12x32}.
After the $B_1$ decay ($1/2\to 1/2+0$) the density matrix is transformed into 
\begin{equation}
\rho_{1/2,\overline{3/2}}^{(f)}=\frac{1}{2}
\sum_{\mu=0}^3\sum_{\bar\nu=0}^{15}C_{\mu\bar\nu}\left(\sum_{\kappa=0}^3 a_{\mu\kappa}\sigma_\kappa^{d}\right)\otimes Q_{\bar\nu}\, ,
\end{equation}
where the $\sigma_k^{d}$ matrices act in the daughter helicity frame.
Correspondingly after the $\bar B_2$ decay ($\overline{3/2}\to \overline{1/2}+0 $) the density matrix would read: 
\begin{equation}
\rho_{1/2,\overline{1/2}}^{(f)}=\frac{1}{2}
\sum_{\mu=0}^3\sum_{\bar\nu=0}^{15}C_{\mu\bar\nu}\,\sigma_\mu\otimes \left(\sum_{\kappa=0}^{3} b_{\bar\nu\kappa} 
\sigma_{\kappa}^d\right)\, .
\end{equation}

Below we provide the explicit expression for the decay matrices $a_{\mu\nu}$ and $b_{\mu\nu}$. Consider a $J=1/2$ or $J=3/2$ hyperon (with initial helicity $\kappa$) decaying into a $J=1/2$ baryon (with helicity $\lambda_1=\lambda$) and a pseudoscalar particle ($\lambda_2=0$). 
By evaluating the transition operator between the initial hyperon and the 
daughter baryon state one gets: 
\begin{eqnarray}
  && {}_{\rm d}\langle \theta, \phi, \lambda\vert S \vert 0,0,
\kappa\rangle_{\rm m} =  \nonumber 
{}_{\rm d}\langle \theta, \phi, \lambda \vert J,\lambda \rangle_{\rm d}
  \nonumber \\
&&  \times 
  {}_{\rm d}\langle J,\lambda \vert S \vert J,\kappa\rangle_{\rm m} 
  \nonumber  \times
  {}_{\rm m}\langle J,\kappa \vert 0,0,\kappa\rangle_{\rm m}   \,
  \label{eq:decay1}
\end{eqnarray}
where the angles $\theta$ and $\phi$ are given
with respect to the helicity frame of the mother hyperon ${\rm m}$.
The amplitude $B_{\lambda}=
 {}_{\rm d}\langle J,\lambda\vert S \vert J,\kappa \rangle_{\rm m} $
depends only on
the helicity of the daughter baryon and it is therefore called helicity amplitude. Recalling also Eq.~\eqref{eq:ang-JM} the transition amplitude becomes:
\begin{equation}
{}_{\rm d}\langle \theta, \phi, \lambda\vert S \vert 0,0,\kappa\rangle
_{\rm m}
\propto {\cal D}^{J *}_{\kappa,\lambda}(\Omega) B_{\lambda},
\label{eqn:amphdecay}
\end{equation}
where ${\cal D}^{J *}_{\kappa,\lambda}(\Omega)={\cal D}^{J *}_{\kappa,\lambda}(\phi,\theta,0)$.
The coefficients  $a_{\mu\nu}$ are then obtained by multiplying the amplitude above
by its conjugate and inserting basis $\sigma$ matrices for the mother 
and the daughter baryon:
\begin{equation}
\begin{split}
{a}_{\mu\nu}&=\frac{1}{4\pi} \sum_{\lambda,\lambda'=-1/2}^{1/2}B_{\lambda}B^*_{\lambda'}\times\\
&\sum_{\kappa,\kappa'=-1/2}^{1/2}
(\sigma_\mu)^{\kappa,\kappa'}(\sigma_\nu)^{\lambda',\lambda}{\cal D}^{1/2 *}_{\kappa,\lambda}(\Omega)
{\cal D}^{1/2}_{\kappa',\lambda'}(\Omega). 
\end{split}
\end{equation} 
These coefficients can be rewritten  in terms of the 
decay parameters $\alpha_D$ and $\phi_D$ defined in Ref.~\cite{PDG}. For completeness 
we first relate the helicity amplitudes to the $S$ and $P$
wave amplitudes $A_S$ and $A_P$, corresponding respectively to the 
parity violating and parity conserving transitions. 
If a hyperon of spin $J$ decays (weakly) into a hyperon of spin $S$ and a (pseudo)scalar state, then the relation between 
helicity amplitudes and canonical amplitudes is given by \cite{Jacob:1959at}
\begin{eqnarray}
  \label{eq:hel-LS}
  B_\lambda = \sum\limits_L \left(\frac{2L+1}{2J+1}\right)^{1/2} \, (L,0;S,\lambda\vert J,\lambda) \, A_L 
\end{eqnarray}
where $(s_1,m_1,s_2,m_2\vert s, m)$ is a Clebsch-Gordan coefficient.
For $J=S=1/2$ the helicity amplitudes are\footnote{Note that the Particle Data Group \cite{PDG} uses $-A_P=A_P^{\rm PDG}$.} 
\begin{eqnarray}
B_{-1/2}&=&\frac{A_S+A_P}{\sqrt{2}} \,, \nonumber \\
B_{1/2}&=&\frac{A_S-A_P}{\sqrt{2}} \,.
\label{eqn:JWdec}
\end{eqnarray} 
Using the normalization $|A_S|^2+|A_P|^2=|B_{-1/2}|^2+|B_{1/2}|^2=1$,
the relation between helicity amplitudes and the decay parameters
is:
\begin{eqnarray}
\alpha_D&=&-2\Re(A_S^*A_P)=|B_{1/2}|^2-|B_{-1/2}|^2  \,, \nonumber\\
\beta_D&=&-2\Im(A_S^*A_P)=2\Im(B_{1/2}B_{-1/2}^*)  \,,\label{eqn:dparam} \\
\gamma_D&=&|A_S|^2-|A_P|^2=2\Re(B_{1/2}B_{-1/2}^*)\nonumber   
\end{eqnarray}
where $\beta_D=\sqrt{1-\alpha_D^2}\sin\phi_D$ and 
 $\gamma_D=\sqrt{1-\alpha_D^2}\cos\phi_D$.
The non-zero elements  of the decay matrix $a_{\mu\nu}$  are (where an overall $\frac{1}{4\pi}$ factor is omitted): 
\begin{eqnarray}
a_{00}&=&1 \,, \nonumber\\
a_{03}&=&\alpha_D \,, \nonumber\\
a_{10}&=&\alpha_D\cos\phi\sin\theta \,, \nonumber\\
a_{11}&=&\gamma_D \cos\theta\cos\phi-\beta_D \sin\phi \,, \nonumber\\
a_{12}&=&-\beta_D \cos\theta
\cos\phi-\gamma_D \sin\phi \,, \nonumber\\
a_{13}&=&\sin\theta \cos\phi \,, \nonumber\\
a_{20}&=&\alpha_D\sin\theta \sin\phi \,, \label{eqn:matrixa}\\
a_{21}&=&\beta_D \cos\phi+\gamma_D \cos\theta \sin\phi \,, \nonumber\\
a_{22}&=&\gamma_D\cos\phi-\beta_D \cos\theta \sin\phi \,, \nonumber\\
a_{23}&=&\sin\theta \sin\phi \,, \nonumber\\
a_{30}&=&\alpha_D\cos\theta \,, \nonumber\\
a_{31}&=&-\gamma_D\sin\theta \,, \nonumber\\
a_{32}&=&\beta_D\sin\theta \,, \nonumber\\
a_{33}&=&\cos\theta\, .\nonumber
\end{eqnarray}
Analogously, the elements of the $b_{\mu\nu}$ matrix are given by
\begin{equation}
\begin{split}
{b}_{\mu\nu}&= \frac{1}{2} \sum_{\lambda,\lambda'=-1/2}^{1/2}B_{\lambda}B^*_{\lambda'}\\
&\times\sum_{\kappa,\kappa'=-3/2}^{3/2}
(Q_\mu)^{\kappa,\kappa'}(\sigma_\nu)^{\lambda',\lambda}{\cal D}^{3/2 *}_{\kappa,\lambda}(\Omega)
{\cal D}^{3/2}_{\kappa',\lambda'}(\Omega).\label{eqn:bij} 
\end{split}
\end{equation}
Out of 64 ${b}_{\mu\nu}$ coefficients 12 are zero. { The coefficients
relevant for the inclusive distributions are presented in Eq.~\eqref{eqn:matrixb} as a part of an example in section~\ref{sec:example}. The remaining coefficients are straightforward to obtain.}
As before, we first rewrite the helicity amplitudes in terms of the canonical amplitudes using Eq.~\eqref{eq:hel-LS}:
\begin{eqnarray}
B_{-1/2}&=&\frac{A_P+A_D}{\sqrt{2}}  \,, \nonumber \\
B_{1/2}&=&\frac{A_P-A_D}{\sqrt{2}}  \,.
\end{eqnarray}
In this case the $P$ and $D$ amplitudes $A_P$ and $A_D$ are the contributing ones.
The definition of the decay parameters $\alpha_D$, $\beta_D$ and $\gamma_D$
 is analogous to that of Eq.~\eqref{eqn:dparam}:
\begin{eqnarray}
\alpha_D&=&-2\Re(A_P^*A_D)=|B_{1/2}|^2-|B_{-1/2}|^2  \,, \nonumber\\
\beta_D&=&-2\Im(A_P^*A_D)=2\Im(B_{1/2}B_{-1/2}^*)  \,, \\
\gamma_D&=&|A_P|^2-|A_D|^2=2\Re(B_{1/2}B_{-1/2}^*)  \,. \nonumber \label{eqn:dparam2}
\end{eqnarray}
Again, they can be expressed in terms of the 
  parameters $\alpha_D$ and $\phi_D$.

\section{Examples}
\label{sec:example}
We discuss the same examples as in Ref.~\cite{Chen:2007zzf} with the aim to
to provide the correct
expressions for reference in ongoing experimental analyses and to
illustrate  how to apply our modular method.
  In particular the discussed reactions could provide 
  an independent verification of 
the new $\Lambda\to p\pi^-$ decay asymmetry parameter value from BESIII.

\subsection{$e^+e^-\to J/\psi,\psi(2S)\to \Lambda\bar\Lambda$}\label{sub:A}
This example is a verification of the angular distributions
derived in \cite{Faldt:2017kgy} and used in the BESIII analysis \cite{Ablikim:2018zay}. We start from  the two-particle
density matrix for  the $\Lambda$-$\bar\Lambda$ pair coming from the $e^+e^-\to
\Lambda\bar \Lambda$ reaction, which is given by Eq.~\eqref{eqn:sig12}.
After considering the 
subsequent two-body weak decays into $p\pi^-/\bar{p}\pi^+$,
the joint angular distribution of the $p/\bar{p}$ pair
is given within the present formalism as:
\begin{equation}
  \mathrm{Tr}\rho_{p\bar{p}}\propto\sum_{\mu,\bar\nu=0}^3C_{\mu\bar\nu}(\theta_\Lambda)\,
  a_{\mu0}^{\Lambda}
{a}_{\bar\nu0}^{\bar\Lambda} \, ,
\label{eq:finsV}
\end{equation}
{with the $a_{\mu0}$ matrices given by Eq.~\eqref{eqn:matrixa}:
$a_{\mu0}^\Lambda\to a_{\mu0}^\Lambda(\theta_p,\phi_p;\alpha_\Lambda)$ and
$a_{\bar\nu0}^{\bar\Lambda}\to a_{\bar\nu0}^{\bar\Lambda}(\theta_{\bar
  p},\phi_{\bar p};\alpha_{\bar\Lambda})$, where only the decay asymmetries
$(\alpha_\Lambda=\alpha_-)/(\alpha_{\bar\Lambda}=\alpha_+)$ for $(\Lambda\to
p\pi^-)/(\bar\Lambda\to\bar{p}\pi^+)$ enter.  The variables $\theta_p$ and
$\phi_p$ are the proton spherical coordinates
in the $\Lambda$ helicity frame with
the axes ${\bf\hat x}_1,{\bf\hat y}_1,{\bf \hat z}_1$ defined in
Fig.~\ref{fig:axes}. The variables $\theta_{\bar p}$ and $\phi_{\bar p}$ are the
antiproton spherical angles in the $\bar\Lambda$ helicity frame with the axes
${\bf\hat x}_2,{\bf\hat y}_2,{\bf\hat z}_2$.}

{The resulting joint angular distribution fully agrees with the
  covariant calculations of Ref.~\cite{Faldt:2017kgy}. In order to
  compare the results one should take into account the different
  definitions of the axes.  The $\Lambda$ scattering angle, $\theta$,
  is defined in Ref.~\cite{Faldt:2017kgy} with respect to the $e^-$
  beam direction ($-{\bf\hat z}$ direction in Fig.~\ref{fig:axes})
  and therefore $\theta=\pi-\theta_\Lambda$.  In addition
  Ref.~\cite{Faldt:2017kgy} uses a common orientation of
  the coordinate systems to represent both
  proton and antiproton directions in the $\Lambda$ and $\bar\Lambda$
  rest frames, respectively. The orientation of this reference
  system can be expressed by the orientations of the helicity
  frames used in this Report as: $(-{\bf\hat x}_1,-{\bf\hat y}_1,{\bf
    \hat z}_1)\equiv (-{\bf\hat x}_2,{\bf\hat y}_2,-{\bf\hat z}_2)$. }

\subsection{$e^+e^-\to J/\psi,\psi(2S)\to \Sigma^0\bar\Sigma^0$}\label{sub:B}

Here we discuss exclusive decay chain: $e^+e^-\to J/\psi,\psi(2S)\to \Sigma^0\bar\Sigma^0$ where $\Sigma^0(\bar\Sigma^0)$ decays electromagnetically
$\Sigma(\bar\Sigma^0)\to\Lambda(\bar\Lambda)\gamma$ and then $\Lambda(\bar\Lambda)$ decays weakly: $\Lambda\to p\pi^-(\bar\Lambda\to \bar p\pi^+)$. In Ref.~\cite{Faldt:2017yqt} it was shown that
the electromagnetic part of the decay chain, where the photon
polarization is not measured, could be  represented by
 decay matrix
as $\check a_{\mu\nu}$ where the only non-zero terms are:
\begin{eqnarray}
\check a_{00}&=&1 \,, \nonumber\\
\check a_{13}&=&-\sin\theta \cos\phi \,, \nonumber\\
\check a_{23}&=&-\sin\theta \sin\phi \,, \label{eqn:matrixag}\\
\check a_{33}&=&-\cos\theta\, ,\nonumber
\end{eqnarray}
where for $\Sigma^0\to\Lambda\gamma(\bar\Sigma^0\to\bar\Lambda\gamma)$
the spherical coordinates $\theta$ and $\phi$ of the daughter
$\Lambda(\bar\Lambda)$ momentum are given in the $\Sigma^0(\bar\Sigma^0)$
helicity frame. The $\check a_{\mu\nu}$ matrix does not involve any decay
parameters and therefore it is only a function of the spherical
coordinates -- $\check a_{\mu\nu}(\theta,\phi)$.  The two body spin
density matrix for the produced $\Sigma^0\bar\Sigma^0$ is given by
Eq.~\eqref{eqn:sig12}.  After including the sequential decays using
our prescription and taking trace of the final proton-antiproton spin
density matrix one has:
\begin{equation}
\mathrm{Tr}\rho_{p\bar{p}}\propto \sum_{\mu,\bar\nu=0}^3\sum_{\mu'=0}^3\sum_{\bar\nu'=0}^3C_{\mu\bar\nu}\ 
    \check a_{\mu\mu'}^{\Sigma^0}\ a_{\mu'0}^\Lambda\ 
    \check a_{\bar\nu\bar\nu'}^{\bar\Sigma^0}\ a_{\bar\nu'0}^{\bar\Lambda}\ ,
\end{equation}
where the $a_{\mu\nu}$ matrices for $1/2\to 1/2+0$ decays are given by Eq.~\eqref{eqn:matrixa} and $\check a_{\mu\mu'}^{\Sigma^0}\to\check  a_{\mu\mu'}^{\Sigma^0}(\theta_\Lambda,\phi_\Lambda)$,
$\check a_{\bar\nu\bar\nu'}^{\bar\Sigma^0}\to \check a_{\bar\nu\bar\nu'}^{\bar\Sigma^0}(\theta_{\bar\Lambda},\phi_{\bar\Lambda})$.

\subsection{$e^+e^-\to J/\psi,\psi(2S)\to \Xi\bar\Xi$}\label{sub:C}

Here we discuss an exclusive decay chain: $e^+e^-\to J/\psi,\psi(2S)\to\Xi\bar\Xi $ where $\Xi(\bar\Xi)$
decays weakly
$\Xi(\bar\Xi)\to\Lambda(\bar\Lambda)\pi$ and then $\Lambda(\bar\Lambda)$ decays weakly: $\Lambda\to p\pi^-(\bar\Lambda\to \bar p\pi^+)$.
The production spin density matrix is given by Eq.~\eqref{eqn:c1212}: $C_{\mu\bar\nu}\to C_{\mu\bar\nu}(\theta_\Xi;\alpha_{\psi},\Delta\Phi)$.
Using replacements Eg.~\eqref{eqn:decay12p} for the sequential decays
and finally taking trace for the unmeasured polarization 
of the final proton-antiproton system
one obtains the differential distribution in the form: 
\begin{equation}
\mathrm{Tr}\rho_{p\bar{p}}\propto \sum_{\mu,\bar\nu=0}^3\sum_{\mu'=0}^3\sum_{\bar\nu'=0}^3C_{\mu\bar\nu}\ 
    a_{\mu\mu'}^\Xi\ a_{\mu'0}^\Lambda\ 
    a_{\bar\nu\bar\nu'}^{\bar\Xi}\ a_{\bar\nu'0}^{\bar\Lambda}\ ,\label{eq:XiXi}
\end{equation}
where the $a_{\mu\nu}$ matrices for $1/2\to 1/2+0$ decays are given by Eq.~\eqref{eqn:matrixa}. The matrices are the functions of the corresponding 
helicity variables and decay parameters: $a_{\mu\mu'}^\Xi\to a_{\mu\mu'}^\Xi(\theta_\Lambda,\phi_\Lambda; \alpha_{\Xi},\beta_\Xi,\gamma_\Xi)$,
 $a_{\bar\nu\bar\nu'}^{\bar\Xi}\to a_{\bar\nu\bar\nu'}^{\bar\Xi}(\theta_{\bar\Lambda},\phi_{\bar\Lambda}; \alpha_{\bar\Xi},\beta_{\bar\Xi},\gamma_{\bar\Xi})$,
$a_{\mu'0}^\Lambda\to a_{\mu'0}^\Lambda(\theta_p,\phi_p; \alpha_{\Lambda})$ and $a_{\bar\nu'0}^{\bar\Lambda}\to a_{\bar\nu'0}^{\bar\Lambda}(\theta_{\bar p},\phi_{\bar p}; \alpha_{\bar\Lambda})$.

With the information provided in this Report -- explicit form of the
matrices $C_{\mu\bar\nu}$ (Eq.~\eqref{eqn:c1212}) and $a_{\mu\nu}$
(Eq.~\eqref{eqn:matrixa}) -- it is straightforward to write a program to
calculate the joint angular distribution using Eq.~\eqref{eq:XiXi}. The result is
much more complicated than given in
Ref.~\cite{Chen:2007zzf}.  We find that from $4^4=256$ possible terms
100 are not equal zero.  An important part of a practical application of
the expression in the maximum likelihood fits to data, such as used in
analysis of Ref.~\cite{Ablikim:2018zay}, is a normalization of the
probability density function using phase space distributed simulated
events
which are processed to include detector and reconstruction effects.
This sample has
to be much larger than data and therefore calculation of the normalization
factor for each parameter set determines
the speed of the fitting procedure. However, Eq.~\eqref{eq:XiXi}
can be rewritten as a polynomial where each term contains
product of a function of decay parameters
and a function of helicity variables:
\begin{equation}
\frac{d\Gamma}{d{\xi}}\propto\mathrm{Tr}\rho_{p\bar{p}}=  \sum_{i=1}^N f_i(\boldsymbol{\pi})\cdot {\cal T}_i(\boldsymbol{\xi})\ , 
\end{equation}
 where $\boldsymbol{\pi}$ represents all the parameters describing the production reaction and the decays
$\boldsymbol{\pi}= (\alpha_{\psi},\Delta\Phi,
  \alpha_{\Xi},\beta_\Xi,\gamma_\Xi,\alpha_{\bar\Xi},\beta_{\bar\Xi},\gamma_{\bar\Xi},
  \alpha_{\Lambda},  \alpha_{\bar\Lambda})$,  $\boldsymbol{\xi}$ represents the full set of nine helicity angles:
$\boldsymbol{\xi}=(\theta_\Xi,\theta_\Lambda,\phi_\Lambda,\theta_p,\phi_p,\theta_{\bar\Lambda},\phi_{\bar\Lambda},\theta_{\bar p},\phi_{\bar p})$ to specify an event and $d\xi$ is the corresponding multidimensional volume element of the phase space
parameterized by the set $\boldsymbol{\xi}$ of the
helicity angles.
 Such representation allows to pre-calculate the normalization integral as:
\begin{equation}
  \int \left(\frac{d\Gamma}{d{\xi}}\right)\epsilon(\boldsymbol{\xi}) d{\xi}
  =  \sum_{i=1}^N f_i(\boldsymbol{\pi})\cdot {\cal I}_i\ , 
\end{equation}
where $\epsilon(\boldsymbol{\xi})$ is multidimensional acceptance-efficiency. The integrals:
\begin{equation}
  {\cal I}_i=\int{\cal T}_i(\boldsymbol{\xi})\epsilon(\boldsymbol{\xi}) d{\xi}
\end{equation}
are independent of the fitted parameters and therefore do not need to
be evaluated at each minimization step. We have found that $N=72$ such
base functions are needed for this reaction. This procedure allows for
a dramatic speed-up of the minimization, what is of importance for the
data sets of several hundreds of thousands fully reconstructed events
as available at BESIII. The same technique can be applied to all other
sequential decays discussed in this Report.

\subsection{$e^+e^-\to \psi(2S)\to \Omega^-\bar\Omega^+$}\label{sub:D}
The expression for two particle spin density matrix  for the $3/2+\overline{3/2}$
final state
is given by Eq.~\eqref{eqn:dens32x32}. Having in mind practical application to BESIII
data we focus on the inclusive reaction, where only the decay products of the $\Omega^-$ are measured.  
In this example the $\Omega^-$ produced in the $e^+e^-\to
\Omega^-\bar\Omega^+$ reaction is identified using the following  sequence of decays: (a)
$\Omega^-\to \Lambda K^-$ and (b) $\Lambda\to p\pi^-$.  To describe
the decay chain we introduce helicity reference frames and the
spherical coordinates $(\theta_\Lambda,\phi_\Lambda)$ and
$(\theta_p,\phi_p)$ for the $\Lambda$ and $p$ directions,
respectively. The scattering
angle of $\Omega$ in the overall CM system is denoted as
$\theta_\Omega$. 
The density matrix of $\Omega^-$ is given by
Eq.~\eqref{eqn:dens32don}:
\[
\rho_\Omega=\sum_{\mu=0}^{15}r_\mu(\theta_\Omega;h_1,h_2,h_3,h_4) \, Q_\mu\, ,
\]
where only seven real coefficients $r_\mu$ are
non-zero and are given by Eq.~\eqref{eqn:r32}.
The $r_\mu$ parameters depend on the scattering
angle $\theta_\Omega$ 
and on four complex form factors. If we are not interested
in the overall normalization then only 
six real parameters are enough to describe the $\Omega$ production process. 
They have to be determined by fitting to the experimental data.  
  The density matrix of the 
$\Lambda$ coming from the $\Omega^-\to \Lambda K^-$ decay can be obtained from Eq.~\eqref{eq:rhoDaughter}:
\[
\rho_\Lambda=\sum_{\mu=0}^{15}\sum_{\nu=0}^3r_\mu\cdot
b_{\mu\nu}^\Omega(\theta_\Lambda,\phi_\Lambda;\alpha_\Omega,\beta_\Omega,\gamma_\Omega)
\sigma_\nu^\Lambda,
\]
where the $b_{\mu\nu}^\Omega$ coefficients depend on the $\Lambda$ angles 
in the $\Omega$ helicity frame
and on the decay parameters of the $\Omega$.
Only 20 of them contribute here, they are given by (where an overall $\frac{1}{8\pi}$ factor is omitted):
\begin{eqnarray}
b_{0,0}^\Omega&=&1 \,, \nonumber\\
b_{1,3}^\Omega&=&\sqrt{\frac{3}{5}}\sin\theta_\Lambda\sin\phi_\Lambda \,, \nonumber\\
b_{6,0}^\Omega&=&-\frac{\sqrt{3}}{4}(3\cos2\theta_\Lambda+1) \,, \nonumber\\
b_{7,0}^\Omega&=&-3\sin\theta_\Lambda\cos\theta_\Lambda\cos\phi_\Lambda \,, \nonumber\\
b_{8,0}^\Omega&=&-\frac{3}{2} \sin^2\theta_\Lambda\cos2\phi_\Lambda \,, \nonumber\\
b_{10,3}^\Omega&=&-9 \sin^2\theta_\Lambda \cos\theta_\Lambda\cos\phi_\Lambda\sin\phi_\Lambda \,, \nonumber\\
b_{11,3}^\Omega&=&-\frac{9}{4\sqrt{10}}(5\cos2\theta+3)\sin\phi_\Lambda\sin\theta_\Lambda  \,, 
\label{eqn:matrixb}\\
b_{0,3}^\Omega&=&\alpha_\Omega b_{0,0}^\Omega \,, \nonumber\\
b_{1,0}^\Omega&=&\alpha_\Omega b_{1,3}^\Omega \,, \nonumber\\
b_{6,3}^\Omega&=&\alpha_\Omega b_{6,0}^\Omega \,, \nonumber\\
b_{7,3}^\Omega&=&\alpha_\Omega b_{7,0}^\Omega \,, \nonumber\\
b_{8,3}^\Omega&=&\alpha_\Omega b_{8,0}^\Omega \,, \nonumber\\
b_{10,0}^\Omega&=&\alpha_\Omega b_{10,3}^\Omega \,, \nonumber\\
b_{11,0}^\Omega&=&\alpha_\Omega b_{11,3}^\Omega \,, \nonumber\\
b_{1,1}^\Omega&=&2\sqrt{\frac{3}{5}}(\beta_\Omega \cos\phi_\Lambda+\gamma_\Omega\cos\theta\sin\phi_\Lambda) \,, \nonumber\\
b_{1,2}^\Omega&=&2\sqrt{\frac{3}{5}}(\gamma_\Omega \cos\phi_\Lambda-\beta_\Omega\cos\theta_\Lambda\sin\phi_\Lambda) \,, \nonumber\\
b_{j,1}^\Omega&=&\gamma_\Omega H_j + \beta_\Omega G_j\, , \, \, \text{for }\, j=10,11, \nonumber\\
b_{j,2}^\Omega&=&\gamma_\Omega G_j -\beta_\Omega H_j \, , \, \, \text{for }\, j=10,11 \nonumber
\end{eqnarray}
where
\begin{eqnarray}
H_{10}&=&-\frac{3}{4}(3\cos2\theta_\Lambda+1)\sin2\phi_\Lambda\sin\theta \,, \nonumber\\
G_{10}&=&-3\sin\theta_\Lambda\cos\theta_\Lambda\cos2\phi_\Lambda \,, \nonumber\\
H_{11}&=&-\frac{3}{8\sqrt{10}}(\cos\theta_\Lambda+15\cos3\theta_\Lambda)\sin\phi_\Lambda \,, \nonumber\\
G_{11}&=&-\frac{3}{4\sqrt{10}}(5\cos2\theta_\Lambda+3)\cos\phi_\Lambda\nonumber\, .
\end{eqnarray}
Finally including also the last decay of the chain,  $\Lambda\to p\pi^-$, 
the proton density matrix in the proton helicity frame can be obtained:
\[
\begin{split}
\rho_p=\sum_{\mu=0}^{15}\sum_{\nu,\kappa=0}^3r_\mu\cdot
b_{\mu\nu}^\Omega\cdot a_{\nu\kappa}^\Lambda(\theta_p,\phi_p;\alpha_\Lambda,\beta_\Lambda,\gamma_\Lambda)
\sigma_\kappa^p.
\end{split}
\] 
Since the proton polarization 
is not measured, we are only interested in the trace of the
density matrix $\mathrm{Tr}\rho_p$ which gives the differential distribution of the
final state specified by the five kinematic variables 
$\cos\theta_\Omega, \cos\theta_\Lambda
  ,\cos\theta_p,\phi_\Lambda,\phi_p$:
\begin{equation*}
  \mathrm{Tr}\rho_{p}\propto 2
\sum_{\mu=0}^{15}\sum_{\nu=0}^{3}r_\mu b_{\mu\nu}^\Omega a_{\nu 0}^\Lambda,
\end{equation*}
where the relevant $a_{\nu,0}^\Lambda$ can be directly taken from Eq.~\eqref{eqn:matrixa}.

\section{Form factors and helicity amplitudes}
\label{sec:ff}

We follow the definitions of \cite{Korner:1976hv} for constraint-free form factors. When relating them to the helicity amplitudes
we use the conventions of \cite{Jacob:1959at}. This makes our helicity amplitudes $A_{\lambda_1,  \lambda_2}$ somewhat different
from the expressions $\Gamma^{\lambda^* ,\lambda}$ of \cite{Korner:1976hv}. 

The form factors for a particle-antiparticle pair of spin 1/2 and mass $m$ are introduced by
\begin{equation}
  \langle B(p_2,\lambda_2) \, \bar B(p_1,\lambda_1) \vert j_\mu(0) \vert 0 \rangle
  = \bar u(p_2,\lambda_2) \, \Gamma_\mu \, v(p_1,\lambda_1)
  \label{eq:elasticFF12}  
\end{equation}
with the electromagnetic current
\begin{eqnarray}
  \label{eq:defemcurrent}
  j_\mu = \frac23 \, \bar u \gamma_\mu u - \frac13 \, \bar d \gamma_\mu d - \frac13 \, \bar s \gamma_\mu s + \ldots \,. 
\end{eqnarray}
and \cite{Korner:1976hv}
\begin{eqnarray}
  \label{eq:elasticFF12Gam}
  \Gamma_\mu := F_1(q^2) \, \gamma_\mu + F_2(q^2) \, \frac{i \sigma_{\mu\nu} q^\nu}{2m}
\end{eqnarray}
where $q=p_1+p_2$ denotes the momentum of the virtual photon.

These form factors are related to the helicity amplitudes by
\begin{eqnarray}
  A_{+1/2,+1/2} & = & 2 m \left(F_1 + \tau \, F_2 \right)\,, \nonumber \\
  A_{+1/2,-1/2} & = & \sqrt{2q^2} \, \left(F_1 + F_2 \right)\,,
  \label{eq:relelFF12helamp}  
\end{eqnarray}
where $\tau=\frac{q^2}{4m^2}$. We have defined
\begin{eqnarray}
  \label{eq:def-helampl-emcur}
  A_{\lambda_1,\lambda_2} := \sqrt{\frac{3}{4\pi}} \, \langle J=1,M;\lambda_1,\lambda_2 \vert \, j(M) \, \vert 0 \rangle
\end{eqnarray}
with 
\begin{eqnarray}
  j(M=+1) &:=& -\frac{1}{\sqrt{2}} \, (j^1 + i j^2) \,, \nonumber \\ 
  j(M=0) &:=& j^3 \,,  
  \label{eq:defjM} \\ 
  j(M=-1) &:= &\frac{1}{\sqrt{2}} \, (j^1 - i j^2) \,.  \nonumber 
\end{eqnarray}
In the following we will stick to the more compact notation for the
helicity form factors from Section \ref{sec:prod}: $A_{1/2,1/2}=:\F_1$,
$A_{1/2,-1/2}=:\F_2$ etc.
Close to threshold $\tau \approx 1$ one finds 
\begin{eqnarray}
  \label{eq:thres1212}
  \F_1 \approx \frac{1}{\sqrt{2}} \, \F_2  \,.
\end{eqnarray}

The form factors for a particle-antiparticle pair of spin 3/2 and mass $m$ are given by
\begin{eqnarray}
\begin{split}
  &\langle B'(p_2,\lambda_2) \, \bar B'(p_1,\lambda_1) \vert J_\mu(0) \vert 0 \rangle  \\
  &= \bar u^\alpha(p_2,\lambda_2) \, \Gamma_{\alpha\beta\mu} \, v^\beta(p_1,\lambda_1)
  \label{eq:elasticFF32}
\end{split}  
\end{eqnarray}
with \cite{Korner:1976hv}
\begin{eqnarray}
  \Gamma_{\alpha\beta\mu} & := & g_{\alpha\beta} \left( F_1(q^2) \, \gamma_\mu + F_2(q^2) \, \frac{i \sigma_{\mu\nu} q^\nu}{2m} \right) 
  \nonumber \\  && 
  {}+ \frac{q_\alpha \, q_\beta}{m^2} \left( F_3(q^2) \, \gamma_\mu + F_4(q^2) \, \frac{i \sigma_{\mu\nu} q^\nu}{2m} \right) \,.
  \phantom{mm}
  \label{eq:elasticFF32Gam}
\end{eqnarray}
These form factors are related to the helicity amplitudes by
\begin{eqnarray}
  \F_1 & = & 2 m \left(1-\frac43 \, \tau\right) \left(F_1 + \tau \, F_2 \right) \nonumber \\ 
  && {} + 2 m \, \frac43 \, \tau \left(1- \tau \right) \left(F_3 + \tau \, F_4 \right) 
\,, \nonumber \\
  \F_2 & = & - \frac23 \, \sqrt{2q^2} \, \left[-\left(1-2 \, \tau\right) \left(F_1 + F_2 \right)
    \right. \nonumber \\ && \phantom{mmmmm} \left. {}
    - 2 \, \tau \left(1-\tau\right) \left(F_3 + F_4 \right) \right] \,, \nonumber \\
  \F_3 & = & \sqrt{\frac23} \, \sqrt{q^2}  \left(F_1 + F_2 \right) \,, \nonumber \\
  \F_4 & = & 2 m \left(F_1 + \tau \, F_2 \right) \,.
  \label{eq:relelFF32helamp}  
\end{eqnarray}
Close to threshold $\tau \approx 1$ one finds 
\begin{eqnarray}
  \label{eq:thres3232}
\begin{split}
  \F_4 \approx  -3 \, \F_1 \approx \sqrt{\frac32} \, \F_3
 \approx  -\frac{3}{\sqrt{8}} \, \F_2 \,.
\end{split}
\end{eqnarray}

Transition form factors for a particle with $J^P=\frac{3}{2}^+$, mass $M$ and an antiparticle with $J^P=\frac{1}{2}^-$, 
mass $m$ are encoded in
\begin{eqnarray}
  \langle B'(p_2,\lambda_2) \, \bar B(p_1,\lambda_1) \vert J_\mu(0) \vert 0 \rangle  \nonumber \\ 
  = \bar u^\nu(p_2,\lambda_2) \, \Gamma_{\nu\mu} \, v(p_1,\lambda_1)  
  \label{eq:transFF}  
\end{eqnarray}
with \cite{Korner:1976hv}
\begin{eqnarray}
  \Gamma_{\nu\mu} & := & G_1(q^2) \left(q^\nu \gamma^\mu - \slashed{q} \, g^{\nu\mu} \right) \gamma_5 \nonumber \\ && {}
  + G_2(q^2) \left(q^\nu p_2^\mu - (q \cdot p_2) \, g^{\nu\mu} \right) \, \gamma_5 \nonumber \\ && {}
  + G_3(q^2) \left(q^\nu q^\mu - q^2 \, g^{\nu\mu} \right) \gamma_5   \,.
  \label{eq:transFFGam}
\end{eqnarray}
These form factors are related to the helicity amplitudes by
\begin{eqnarray}
  \F_1 & = & \sqrt{\frac23} \, N \, \sqrt{q^2} 
  \left( G_1 + M \, G_2 + \frac{q^2+M^2-m^2}{2 M} \, G_3 \right)  , \nonumber \\
  \F_2 & = & \frac1{\sqrt 3} \, N
  \left( \frac{q^2-m \, (m+M)}{M} \, G_1 
    \right. \nonumber \\ && \phantom{mmmm} \left. {}
    + \frac{q^2+M^2-m^2}{2} \, G_2 + q^2 \, G_3 \right)  \,, 
  \label{eq:reltransFFhelamp}  \\
  \F_3 & = & N \left( (m+M) \, G_1 + \frac{q^2+M^2-m^2}{2} \, G_2 + q^2 \, G_3 \right) \nonumber 
\end{eqnarray}
with a ``normalization factor''
\begin{eqnarray}
  \label{eq:defnormtrans}
  N(q^2) := \sqrt{q^2-(M-m)^2}  \,.
\end{eqnarray}
Close to threshold, $q^2 \approx (m+M)^2$, one finds 
\begin{eqnarray}
  \label{eq:thres3212}
  \F_1 \approx \sqrt{2} \, \F_2 \approx \sqrt{\frac23} \, \F_3 \,.
\end{eqnarray}

To facilitate the matching between Feynman matrix elements and expressions in the helicity framework of 
Jacob and Wick \cite{Jacob:1959at} we provide in appendix \ref{sec:convent} some explicit formulas for the particle and 
antiparticle spinors.

\section{Further discussion}
\label{sec:discussion}

We would like to draw attention to some interesting properties
of the derived angular distributions close to threshold.  
For the production of two spin-1/2 baryons the parameters $\alpha_\psi$ and 
$\Delta\Phi$ are zero at threshold. Therefore, there is no spin polarization 
implying the inclusive
distributions of the decay products are isotropic.  
For the spin $3/2+\overline{3/2}$ production the baryons are polarized even at threshold.
The inclusive distributions
of the decay products would be
isotropic if
  $r_0=1$ (assuming normalization $\left| {\F_2}\right| ^2\!+\!2 (\left|{\F_1}\right| ^2\!+\!\left| {\F_3}\right| ^2\!+\! \left| {\F_4}\right| ^2)=1$) and all
other $r_i$ terms were zero in Eq.~\eqref{eqn:r32}. 
Using the close-to-threshold relation between the form factors from
Eq.~\eqref{eq:thres3232} one sees that three additional terms are not zero:
\begin{eqnarray}
r_6&=&\frac{1}{5 \sqrt{3}}(1-3\cos^2\!\theta_1) \,, \nonumber \\
r_7&=&\frac{1}{5}\sin\!2\theta_1 \,, \label{eq:r678dis} \\
r_8&=&-\frac{1}{5}\sin^2\!\theta_1 \,. \nonumber
\end{eqnarray}
An inclusive distribution that is only differential in the production angle is not sensitive to these parameters. Indeed, 
$\alpha_\psi$ as introduced in Eq.~\eqref{eq:aphapsi32} vanishes at threshold. However, 
distributions differential in the angles of decay products are sensitive. It is not even necessary that the decay is parity 
violating. If one assumed 
that the decay  $3/2\to 1/2+0$ 
would be parity conserving, implying $\gamma_D=1$, then the angular distribution
of the decay products is already not isotropic: 
\[
\frac{d\Gamma}{d\cos\!\theta_1d\cos\!\theta_D}\propto
1+\frac{(1-3\cos^2\!\theta_1)(1-3\cos^2\!\theta_D)}{10}  \,.
\]
This property of the reaction close to threshold could 
be used to establish spin assignment of the produced baryons
by studying inclusive angular distributions.
One possible test is to calculate the 
moment $(1-3\cos^2\!\theta_D)$, where $\theta_D$ is the helicity 
angle of the daughter baryon. For the spin $1/2+\overline{1/2}$
reaction 
this quantity is zero. 

The above  observation could be also expressed using the degree of 
polarization, which is defined for a spin $3/2$ particle as \cite{Doncel:1972ez}:
\begin{equation}
d(3/2)=\sqrt{\sum_{\mu=1}^{15}\left(\frac{r_\mu}{r_0}\right)^2}.
\end{equation} 
It is easy to check
that at threshold  $d(3/2)=\frac{2}{5\sqrt{3}}\approx 23\%$, if the baryon-antibaryon pair
is produced
in an $e^+e^-$ process.

This suggests that the formalism developed here can be used to determine or at least constrain the spin of baryons. 
This is a highly welcome opportunity in view of the fact that only part of the quantum numbers of hyperons have actually been
experimentally confirmed \cite{PDG}. In the present work we have assigned the standard 
properties to the weakly decaying hyperons. To really confirm the quantum numbers one has to calculate the angular
distributions based on 
various spin and parity assignments, compare the results and explore the experimental capabilities to distinguish different
cases. This is beyond the scope of the present work, but constitutes a natural extension of the formalism presented here.

{Coming back to the motivations of our study: we
  provide modular tools to construct joint decay distributions of sequential
  decay processes for the baryon-antibaryon pairs produced at electron
  positron colliders.  Our expressions are specially useful for the
  processes at $J^{PC}=1^{--}$ resonances such as $J/\psi$ and $\psi$
  where the large statistics data sets are available and the contribution from
  the two photon production mechanism is suppressed.  Contrary to the
  previously published calculations using Jacob and  Wick helicity formalism
\cite{Tixier:1988fv,Chen:2007zzf,Ablikim:2009ab}
  we
  find the angular distributions consistent with calculations using Feynman diagrams~\cite{Faldt:2017kgy} for production of a pair of
  spin-$1/2$ baryons. We can reproduce the 
  results of \cite{Tixier:1988fv,Chen:2007zzf,Ablikim:2009ab} by replacing
  the correct density matrix of the virtual photon
  Eq.~\eqref{eq:explicit-init} by its diagonal part. One important
  conclusion is that the two experimental analyses of $J/\psi\to\Lambda\bar\Lambda$ \cite{Tixier:1988fv,Ablikim:2009ab} used not correct joint angular
  distributions and the reported  results for
  $\alpha_+$ should be re-evaluated. Once validated
  for the spin $1/2+\overline{1/2}$ case,
  the  helicity formalism together with the base spin
  matrices~\cite{Doncel:1972ez},
   allows for a straightforward extension to the production of
  higher spin  baryon states.
  Our systematic derivation demonstrates that a special
  care has to be taken to match the 
  definition of the helicity variables with the
  amplitude transformations used.  The presented formalism is applied in a
  computer program to calculate the angular distributions using well defined
  modules for the production and the sequential decays. In particular the derived formulas
  for $e^+e^-\to J/\psi,\psi(2S)\to \Sigma^0\bar\Sigma^0$
  (Sec.~\ref{sub:B}) and $e^+e^-\to J/\psi,\psi(2S)\to \Xi^-\bar\Xi^+$
  (Sec.~\ref{sub:C}) will be used to search for transverse polarization and, if the
  polarization is found, to independently verify the new value for the
  $\alpha_{-}$ parameter.}
\acknowledgments

We would like to thank Changzheng Yuan for initiating this project
and for the support.  We are grateful to Patrik Adlarson
for useful discussions.  AK would like to thank Shuangshi Fang
for support for the visit at IHEP and acknowledges
grant of Chinese
Academy of Science President's
International Fellowship Initiative (PIFI) for Visiting Scientist.
 
\appendix

\section{Spin $\frac{3}{2}$ basis matrices}
\label{sec:mat32}
 To describe a spin-3/2 particle density matrix the following set of
 $Q_M^{L}$ matrices with $0\le L\le 3$ and $-L\le M\le L$ is needed,
 in total 16 $4\times 4$ matrices. The matrices are introduced in 
Ref.~\cite{Doncel:1972ez}.
$Q_0^0=\frac{1}{3}\mathds{1}_4$ where $\mathds{1}_4$ is the identity $4\times 4$
matrix. We use the following notation
with only one index to denote the matrices:
\begin{equation}
Q_{L(L+1)+M} := \frac{3}{4}Q_M^L.
\end{equation} 
Given the index $\mu$ belonging to the matrix 
$Q_\mu$, the  corresponding values of $M$ and $L$
 can be easily retrieved:
\begin{equation}
\begin{array}{cccc}
\mu=0&:&L=0\,,&M=0\,,\\
1\le\mu\le3&:&L=1\,,& -1\le M\le 1\,,\\
4\le\mu\le8&:&L=2\,,& -2\le M\le 2\,,\\
9\le\mu\le15&:&L=3\,,& -3\le M\le 3\, .
\end{array}
\end{equation}

Below the explicit expressions for the $Q_M^{L}$ matrices
are provided:
\begin{eqnarray*}
Q_{-1}^1&=&\frac{i}{\sqrt{5}}\left(
\begin{array}{cccc}
 0 & -1 & 0 & 0 \\
 1 & 0 & -\frac{2}{\sqrt{3}} & 0 \\
 0 & \frac{2}{\sqrt{3}} & 0 & -1 \\
 0 & 0 & 1 & 0 \\
\end{array}
\right) \,, \\
\end{eqnarray*}
\begin{eqnarray*}
Q_0^1&=&\sqrt{\frac{3}{5}}\left(
\begin{array}{cccc}
 1 & 0 & 0 & 0 \\
 0 & \frac{1}{{3}} & 0 & 0 \\
 0 & 0 & -\frac{1}{{3}} & 0 \\
 0 & 0 & 0 & -1 \\
\end{array}
\right) \,, \\
\end{eqnarray*}
\begin{eqnarray*}
Q_{1}^1&=& \frac{1}{\sqrt{5}}\left(
\begin{array}{cccc}
 0 & 1 & 0 & 0 \\
 1 & 0 & \frac{2}{\sqrt{3}} & 0 \\
 0 & \frac{2}{\sqrt{3}} & 0 & 1 \\
 0 & 0 & 1 & 0 \\
\end{array}
\right) \,, \\
\end{eqnarray*}
\begin{eqnarray*}
Q_{-2}^2&=&\frac{i}{\sqrt{3}}\left(
\begin{array}{cccc}
 0 & 0 & -1 & 0 \\
 0 & 0 & 0 & -1 \\
 1 & 0 & 0 & 0 \\
 0 & 1 & 0 & 0 \\
\end{array}
\right) \,, \\
\end{eqnarray*}
\begin{eqnarray*}
Q_{-1}^2&=&\frac{i}{\sqrt{3}}\left(
\begin{array}{cccc}
 0 & -1 & 0 & 0 \\
 1 & 0 & 0 & 0 \\
 0 & 0 & 0 & 1 \\
 0 & 0 & -1 & 0 \\
\end{array}
\right) \,, \\
\end{eqnarray*}
\begin{eqnarray*}
Q_{0}^2&=&\frac{1}{\sqrt{3}}\left(
\begin{array}{cccc}
 1 & 0 & 0 & 0 \\
 0 & -1 & 0 & 0 \\
 0 & 0 & -1 & 0 \\
 0 & 0 & 0 & 1 \\
\end{array}
\right) \,, \\
\end{eqnarray*}
\begin{eqnarray*}
Q_{1}^2&=&\frac{1}{\sqrt{3}}\left(
\begin{array}{cccc}
 0 & 1 & 0 & 0 \\
 1 & 0 & 0 & 0 \\
 0 & 0 & 0 & -1 \\
 0 & 0 & -1 & 0 \\
\end{array}
\right) \,, \\
\end{eqnarray*}
\begin{eqnarray*}
Q_{2}^2&=&\frac{1}{\sqrt{3}}\left(
\begin{array}{cccc}
 0 & 0 & 1 & 0 \\
 0 & 0 & 0 & 1 \\
 1 & 0 & 0 & 0 \\
 0 & 1 & 0 & 0 \\
\end{array}
\right) \,, \\
\end{eqnarray*}
\begin{eqnarray*}
Q_{-3}^3&=&i \sqrt{\frac{2}{3}}\left(
\begin{array}{cccc}
 0 & 0 & 0 & -1 \\
 0 & 0 & 0 & 0 \\
 0 & 0 & 0 & 0 \\
 1 & 0 & 0 & 0 \\
\end{array}
\right) \,, \\
\end{eqnarray*}
\begin{eqnarray*}
Q_{-2}^3&=&\frac{i}{\sqrt{3}}\left(
\begin{array}{cccc}
 0 & 0 & -1 & 0 \\
 0 & 0 & 0 & 1 \\
 1 & 0 & 0 & 0 \\
 0 & -1 & 0 & 0 \\
\end{array}
\right) \,, \\
\end{eqnarray*}
\begin{eqnarray*}
Q_{-1}^3&=&i
\sqrt{\frac{2}{5}}\left(
\begin{array}{cccc}
 0 & -\sqrt{\frac{1}{3}} & 0 & 0 \\
 \sqrt{\frac{1}{3}} & 0 & 1  & 0 \\
 0 & -1 & 0 & -\sqrt{\frac{1}{3}} \\
 0 & 0 & \sqrt{\frac{1}{3}} & 0 \\
\end{array}
\right) \,, \\
\end{eqnarray*}
\begin{eqnarray*}
Q_{0}^3&=&\sqrt{\frac{3}{5}}\left(
\begin{array}{cccc}
 \frac{1}{3} & 0 & 0 & 0 \\
 0 & -1 & 0 & 0 \\
 0 & 0 & 1 & 0 \\
 0 & 0 & 0 & -\frac{1}{3} \\
\end{array}
\right) \,, \\
\end{eqnarray*}
\begin{eqnarray*}
Q_{1}^3&=&\sqrt{\frac{2}{5}}\left(
\begin{array}{cccc}
 0 & \sqrt{\frac{1}{3}} & 0 & 0 \\
 \sqrt{\frac{1}{3}} & 0 & -1 & 0 \\
 0 & -1 & 0 & \sqrt{\frac{1}{3}} \\
 0 & 0 & \sqrt{\frac{1}{3}} & 0 \\
\end{array}
\right) \,, \\
\end{eqnarray*}
\begin{eqnarray*}
Q_{2}^3&=&\frac{1}{\sqrt{3}}\left(
\begin{array}{cccc}
 0 & 0 & 1 & 0 \\
 0 & 0 & 0 & -1 \\
 1 & 0 & 0 & 0 \\
 0 & -1 & 0 & 0 \\
\end{array}
\right) \,, \\
\end{eqnarray*}
\begin{eqnarray*}
Q_{3}^3&=&\sqrt{\frac{2}{3}}\left(
\begin{array}{cccc}
 0 & 0 & 0 & 1 \\
 0 & 0 & 0 & 0 \\
 0 & 0 & 0 & 0 \\
 1 & 0 & 0 & 0 \\
\end{array}
\right)\, .
\end{eqnarray*}

\section{Conventions for spin-1/2, spin-1 and spin-3/2 spinors for particles and antiparticles}
\label{sec:convent}

Various conventions for spinors are used in the literature. Not all of them fit to the helicity 
framework of Jacob and Wick \cite{Jacob:1959at}. Therefore we provide here 
some explicit formulas for the spinors. To this end one has to be careful in the 
construction of the 
states denoted by type 2 in \cite{Jacob:1959at} as they are not obtained by just a rotation. 
As spelled out in \cite{Jacob:1959at}, two-particle states flying in an arbitrary direction are obtained by two-particle states
where state 1 flies in the $(+z)$ direction and state 2 in the $(-z)$ direction. In the following we present explicitly the 
spinors for the states 1 and 2 with which one starts. We use the Pauli-Dirac representation for the gamma matrices. 
For the spin-1/2 states with helicity $\lambda_{1,2}$, mass $m$, energy $E$, and momenta $p_z$ or $-p_z$ ($p_z \ge 0$) one finds
\begin{equation*}
  \begin{split}  
    u(p_z,\pm 1/2) = &\left(
      \begin{array}{r}
        \sqrt{E+m} \; \chi_\pm \\[0.5em] \pm \sqrt{E-m} \; \chi_\pm
      \end{array}
    \right)  \,, \\[0.5em]
    u(-p_z,\pm 1/2) = &\left(
      \begin{array}{r}
        \sqrt{E+m} \; \chi_\mp \\[0.5em] \pm \sqrt{E-m} \; \chi_\mp
      \end{array}
    \right)  \,, \\[0.5em]
    v(p_z,\pm 1/2) = &\left(
      \begin{array}{r}
        \sqrt{E-m} \; \chi_\mp \\[0.5em] \mp \sqrt{E+m} \; \chi_\mp
      \end{array}
    \right)  \,,  \\[0.5em]
    v(-p_z,\pm 1/2) = &\left(
      \begin{array}{r}
        -\sqrt{E-m} \; \chi_\pm \\[0.5em] \pm \sqrt{E+m} \; \chi_\pm
      \end{array}
    \right) 
  \end{split}
\end{equation*}
with the two-component spinors
\begin{eqnarray*}
  \chi_+ := \left(
    \begin{array}{c}
      1 \\ 0
    \end{array}
  \right) \,, \qquad 
  \chi_- := \left(
    \begin{array}{c}
      0 \\ 1
    \end{array}
  \right)  \,.
\end{eqnarray*}

For the spin-1 states with helicity $\lambda_{1,2}$, mass $m$, energy $E$, and momenta $p_z$ or $-p_z$ ($p_z \ge 0$) we use
\begin{eqnarray*}
  \varepsilon^\mu(p_z,+1) &=& \frac{1}{\sqrt{2}} \left(
      0, -1, -i , 0
  \right)  \,, \\
  \varepsilon^\mu(p_z,0) &=& \frac{1}{m} \left(
      p_z , 0,  0, E
  \right)  \,, 
  \\
  \varepsilon^\mu(p_z,-1) &=& \frac{1}{\sqrt{2}} \left(
      0, 1, -i, 0
  \right)  \,,  
\end{eqnarray*}
\begin{eqnarray*}
  \varepsilon^\mu(-p_z,+1) &=& \frac{1}{\sqrt{2}} \left(
      0, 1, -i , 0
  \right)  \,, \\
  \varepsilon^\mu(-p_z,0) &=& \frac{1}{m} \left(
      -p_z , 0,  0, E
  \right)  \,, 
  \\
  \varepsilon^\mu(-p_z,-1) &=& \frac{1}{\sqrt{2}} \left(
      0, -1, -i, 0
  \right)  \,. 
\end{eqnarray*}

Finally, we present explicit expressions for the spin-3/2 states with helicity $\lambda_{1,2}$, mass $m$, energy $E$, and momenta $p_z$ or $-p_z$ ($p_z \ge 0$):
\begin{eqnarray*}
    u^\mu(p_z,\pm 3/2) &=& \varepsilon^\mu(p_z,\pm 1) \, u(p_z,\pm 1/2)  \,,  \\[0.5em]
    u^\mu(p_z,\pm 1/2) &=& 
    \frac{1}{\sqrt{3}} \, \varepsilon^{\mu}(p_z,\pm 1) \, u(p_z, \mp 1/2) \\
    &&{} + \sqrt{\frac23} \, \varepsilon^{\mu}(p_z,0) \, u(p_z, \pm 1/2) \,,  \\[0.5em]
    u^\mu(-p_z,\pm 3/2) &=&  \varepsilon^\mu(-p_z,\pm 1) \, u(-p_z,\pm 1/2)  \,, \\[0.5em]
    u^\mu(-p_z,\pm 1/2) &=& 
    \frac{1}{\sqrt{3}} \, \varepsilon^{\mu}(-p_z,\pm 1) \, u(-p_z, \mp 1/2) \\
    &&{} + \sqrt{\frac23} \, \varepsilon^{\mu}(-p_z,0) \, u(-p_z, \pm 1/2) \,, \\[0.5em]
    v^\mu(p_z,\pm 3/2) &=& \varepsilon^{\mu*}(p_z,\pm 1) \, v(p_z,\pm 1/2)  \,, \\[0.5em]
    v^\mu(p_z,\pm 1/2) &=& \frac{1}{\sqrt{3}} \, \varepsilon^{\mu*}(p_z,\pm 1) \, v(p_z, \mp 1/2) \\
    &&{} + \sqrt{\frac23} \, \varepsilon^{\mu*}(p_z,0) \, v(p_z, \pm 1/2) \,, \\[0.5em]
    v^\mu(-p_z,\pm 3/2) &=& \varepsilon^{\mu*}(-p_z,\pm 1) \, v(-p_z,\pm 1/2) , \\[0.5em]
    v^\mu(-p_z,\pm 1/2) &=&  
    \frac{1}{\sqrt{3}} \, \varepsilon^{\mu*}(-p_z,\pm 1) \, v(-p_z, \mp 1/2) \\
    &&{} + \sqrt{\frac23} \, \varepsilon^{\mu*}(-p_z,0) \, v(-p_z, \pm 1/2) \,.
\end{eqnarray*}

In general, if one takes a state flying in the $(+z)$ direction and applies to it just a rotation by $\pi$ around the $y$ axis, 
then the result differs from the Jacob/Wick construction by a factor $(-1)^{s_2-\lambda_2}$. Thus for $s_2=1/2$ one picks up a minus 
sign for $\lambda_2 = -1/2$ while for $s_2=3/2$ one picks up a minus sign for $\lambda_2 = +1/2, -3/2$.

\bibliography{refBB}
\end{document}